\documentclass[aps,twocolumn,superscriptaddress,nofootinbib]{revtex4}

\usepackage{latexsym}
\usepackage[normalem]{ulem}
\usepackage{hyperref,amssymb}
\usepackage{url}
\usepackage{color,soul}
\usepackage{graphicx}
\usepackage{verbatim}
\usepackage{multirow}
\usepackage{amsmath}
\newcommand{\beq}{\begin{eqnarray}}
\newcommand{\eeq}{\end{eqnarray}}
\usepackage{mathrsfs}
\usepackage{float,soul}
\usepackage[dvipsnames]{xcolor}
\usepackage{mathtools}
\usepackage{slashed}
\usepackage{physics}	
\usepackage{graphicx}   
\usepackage{epstopdf}
\usepackage{subfigure}  
\usepackage{hyperref}   
\usepackage{bbold}
\usepackage{wasysym}
\usepackage{feynmp}
\def \bs{\textbf}

\usepackage[latin1]{inputenc}
\usepackage{tikz}

\usetikzlibrary{decorations.pathmorphing}
\usetikzlibrary{shapes.misc}
\tikzset{cross/.style={cross out, draw=black, minimum size=8*(#1-\pgflinewidth), inner sep=0pt, outer sep=0pt},
cross/.default={1pt}}

\begin{document}
\title{\Large Anharmonic theory of superconductivity and its applications to emerging quantum materials}
\author{Chandan Setty$^{\oplus,\dagger}$}
\thanks{settychandan@gmail.com}
\affiliation{Department of Physics and Astronomy, Rice Center for Quantum Materials, Rice University, Houston, Texas 77005, USA}
\author{Matteo Baggioli}
\thanks{b.matteo@sjtu.edu.cn}
\affiliation{Wilczek Quantum Center, School of Physics and Astronomy, Shanghai Jiao Tong University, Shanghai 200240, China \& Shanghai Research Center for Quantum Sciences, Shanghai 201315.}
\author{Alessio Zaccone}
\thanks{alessio.zaccone@unimi.it}
\affiliation{Department of Physics ``A. Pontremoli", University of Milan, via Celoria 16, 20133 Milan, Italy.}
\affiliation{Cavendish Laboratory, University of Cambridge, JJ Thomson Avenue, CB30HE Cambridge, U.K.}

\begin{abstract}
The role of anharmonicity on superconductivity has often been disregarded in the past. Recently, it has been recognized that anharmonic decoherence could play a fundamental role in determining the superconducting properties (electron-phonon coupling, critical temperature, etc) of a large class of materials, including systems close to structural soft-mode instabilities, amorphous solids and metals under extreme high-pressure conditions.
Here, we review recent theoretical progress on the role of anharmonic effects, and in particular certain universal properties of anharmonic damping, on superconductivity. Our focus regards the combination of microscopic-agnostic effective theories for bosonic mediators with the well-established BCS theory and Migdal-Eliashberg theory for superconductivity. We discuss in detail the theoretical frameworks, their possible implementation within first-principles methods, and the experimental probes for anharmonic decoherence. Finally, we present several concrete applications to emerging quantum materials, including hydrides, ferroelectrics and systems with charge density wave instabilities.
\end{abstract}

\maketitle

 \tableofcontents

 \section{Introduction: Anharmonicity and Superconductivity}
 \subsection{Historical perspective: anharmonicity and superconductivity} 
 In classic theories of phonon-mediated superconductivity, such as BCS theory and Migdal-Eliashberg theory, phonons were described as harmonic oscillators. In the second half of the 1970s, the discovery of the ``high-$T_c$'' (for that time) superconductivity in Niobium-based alloys which superconduct at temperatures $T>10$K prompted a change in paradigm.

Those materials were rich in structural instabilities often linked to quasi-localized lattice (ionic) vibrations, or coupled lattice-electronic instabilities of the Jahn-Teller type. 
Plakida and co-workers developed early models addressing the influence, and enhancement, of the critical temperature $T_c$ due to highly anharmonic quasi-local vibrations, in a first series of papers~\cite{Vujicic_1979,Vujicic_1981}. In these papers, the quasi-local vibrations (QLV) were modelled as two-level anharmonic wells by means of a pseudospin formalism. Within the Eliashberg formalism, pairing properties containing contributions from these QLVs, distinct from those of standard phonons, are shown to produce a significant enhancement of the superconducting $T_{c}$.
 
Plakida and others \cite{Plakida_1987} modified these models further to understand the high temperature superconductivity in the copper based materials such as La(Y)BaCuO \cite{Bednorz1986}. 
 In this model, the structural instability, again described in the form of anharmonic wells with two levels, occurs due to the rotational motions of the apical oxygen that are located within the layered perovskite structure.
 Within this theory, highly anharmonic motions by the apical oxygens lead to the enhancement of $T_{c}$. These motions are associated with an amplitude of displacement $d$ that is much greater than the mean-squared displacement $\langle u ^{2}\rangle$ of the ions in the lattice. Given that $\lambda$, the electron-phonon coupling, scales as the ionic motions squared, along with the hierarchy of scales  $d^{2}/\langle u ^{2}\rangle \gg 1$, one can justify a significant rise of $T_c$ caused by the soft, localized unstable vibrations. Such an enhancement is also reflected in an effectively stronger electron-phonon coupling with respect to that of the harmonic limit $\lambda_{harm}$, that is: $\lambda/\lambda_{harm} \sim d^{2}/\langle u ^{2}\rangle \gg 1$. Furthermore, the enhancement may also occur via polaron formation, leading to bi-polaronic theories of high-$T_c$ superconductivity \cite{Alexandrov}.
 
 While this enhanced electron-phonon coupling caused by localized soft vibrational modes of the oxygen atoms is irrefutable (and was shown early on in Raman experiments by K. A. M\"{u}ller, Liarokapis, Kaldis and co-workers \cite{Liarokapis1,LiarokapisPRL,Liarokapis2019}), it does not fully explain the interesting phenomenology of cuprate superconductivity in its entirety. These include $d$-wave symmetry of the paired wavefunction, a non-Fermi liquid normal phase and the effects of magnetic correlations. Additionally, these early models do not provide a systematic relationship between $T_c$ and lattice anharmonicity, since they focus on two-level type excitations modelled as pseudospin excitations, thus leaving out all the usual descriptors of lattice anharmonicity (\textit{i.e.} phonon linewidth, Gr\"{u}neisen parameter, etc).
 Finally, these models, while they predict an enhancement of $T_c$ with anharmonic motions, they are unable to predict other regimes where, instead, the anharmonicity is detrimental for the superconductivity and thus causes a reduction of $T_c$.
 
 More recently, anharmonic extensions of conventional superconductivity theory have been proposed in the context of rattling modes in caged thermoelectric-type materials, such as filled skutterudites, $\beta$-pyrochlore oxides and clathrates \cite{Hotta}. A common feature of this material group is the existence of nano-size cages of light atoms. The ion enclosed in the cage, frequently called the guest ion, experiences a highly anharmonic potential and it can perform large amplitude oscillations, referred to as ``rattling''. 
 
 In the context of superconductivity in this class of materials (a known example is superconductivity in the $\beta$-pyrochlore oxides which appears to be enhanced by anharmonicity \cite{Isono}), Oshiba and Hotta \cite{Hotta} developed a theory of superconductivity using the Holstein ``small polaron'' model as a the starting point to treat electron-phonon contributions to the Hamiltonian where large screening effects lead to small polaron radius and large electron-phonon coupling constant. The next step was to apply Migdal-Eliashberg theory to evaluate the $T_c$ as a function of the quartic and sixth-order anharmonic coefficients in the lattice Hamiltonian. The model predicts an enhancement of $T_c$ with increasing anharmonicity followed by a peak or maximum and then a declining regime -- a superconducting dome.
 
 The question of how anharmonicity affects superconductivity at a more fundamental level has however remained unexplored until recently.
 In \cite{Setty2020} (see also \cite{PhysRevB.106.139901}) the effect of anharmonicity of phonons (both acoustic and optical) has been described at the level of BCS theory. For optical phonons, a non-monotonic dependence of $T_c$ on the parameter which  characterizes the lattice anharmonicity, \textit{i.e.} the phonon damping or linewidth, was predicted, with a dome in $T_c$. The enhancement can be explained thanks to dissipation acting to connect bosons at different energy scales that combine coherently to increase the effective electron-phonon coupling and $T_c$. Mathematically, the wave-vector dependence in the propagator is integrated out in the gap equation, and the integration combines high and low energy phonons coherently to enhance the effective electron-phonon coupling and hence the $T_c$. Such a mechanism was previously discussed in the context of proton irradiated samples of La$_{2-x}$Ba$_x$CuO$_4$~\cite{Setty2019}.

Some experimental evidence of the enhancement regime came in the study of filled skutterudite ${\mathrm{LaRu}}_{4}{\mathrm{As}}_{12}$, by using electron irradiation to tune the phonon anharmonicity \cite{Shibauchi}. In practice, electron irradiation was used to suppress certain anharmonic phonon modes by creation of suitable defects. Upon suppressing the anharmonic phonon modes, the $T_c$ was observed to decrease monotonically in the investigated regime. 
 
 The aim of this review is to further explore, on the basis of the available literature, the effects of lattice anharmonicity, especially anharmonic damping, on superconductivity.
 The emerging picture is that phonon anharmonicity \cite{Wei2021} can be viewed as a powerful means to determine significant variations in $T_c$, including both enhancement and suppression.
 This becomes an issue of vital important in all the high-$T_c$ superconducting materials, from the cuprates (for the reasons explained above) to the high-pressure hydrides \cite{Pickett}, where light hydrogen atoms perform huge anharmonic zero-point motions, and where anharmonicity is key to both determine phase stability and superconductivity. In these materials, and more in general in anharmonic crystals \cite{Cowley_1968}, clear guidelines or understanding about these effects are still lacking.
It is hoped that theoretical concepts as embodied in an ``anharmonic theory of superconductivity'' will provide a deeper understanding of electron-phonon superconductivity in materials with non-trivial lattice effects, such as strong anharmonicity, phonon softening and lattice instabilities.

 \subsection{Boson damping mechanisms}
 Different microscopic origins of damping for the bosons involved in the Cooper pairing of electrons can play a role in the superconductivity mechanisms. These include glassy damping due to disorder, damping due to electron-phonon interaction, damping due to phonon-phonon interactions, just to name the most important ones. 
 In real materials with complex chemistry, the interatomic potential is very far from being harmonic, such that the intrinsically large anharmonicity of the lattice dynamics leads to strong damping of phonons due to phonon-phonon processes.
 The effect of anharmonicity on the phonon dispersion curves is twofold. On one hand, the bare phonon frequency gets strongly renormalized (typically, lowered), while on the other hand the lifetime of the phonon also gets reduced (damping).
  
 Anharmonicity arises from the non-harmonic character of the interatomic interactions, although it can also arise from the electron-phonon interaction itself. Here, we shall focus on the lattice anharmonicity (neglecting the contributions from electron-phonon interaction), and on the damping effect (\textit{i.e.}, we take the phonon frequencies in the effective theories as already renormalized).

 \subsubsection{Akhiezer damping of acoustic phonons}
 \label{lolo}
 At finite temperature, the main mechanism for damping of acoustic phonons with long wavelengths is provided by the Akhiezer mechanism, whose form can be derived directly from hydrodynamics (\textit{i.e.} from viscoelasticity) \cite{Chaikin}. From the point of view of elastodynamics, the viscous contribution to the stress, $\sigma_{ij}'$ (which is dissipative, and therefore odd under time reversal) is added to the elastic component of the stress $\sigma^{el}_{ij}$ to make the overall total stress $\sigma_{ij}$. In the context of linear viscoelasticity theory \cite{christensen2003theory}, this is tantamount to assume the so-called Kelvin-Voigt model, which is the most suitable to describe viscoelastic solids (in contrast to the Maxwell model, which better applies to viscoelastic fluids). The corresponding elastodynamical equations reduce then to~\cite{landau2013fluid,landau1986theory,Chaikin}:
\begin{equation}
\rho\, \ddot{u}_{i}=\nabla_j\, \sigma_{ij}+f_i^{ext}(\textbf{r})\,,\qquad \sigma_{ij}= \sigma^{el}_{ij}+\sigma_{ij}'\,, \label{equa1}
\end{equation}
where Latin indices are used to denote spatial components.
The above equation expresses that the internal stress force $\nabla_k \sigma_{ik}$ plus the external force density $f_i^{ext}(\textbf{r})$ is equal to the acceleration of the elastic displacement field $u_i$ times the mass density $\rho$ of the medium (Newton's law). At a linearized level, \textit{i.e.}, for small deformations, the elastic contribution is given as usual by $\sigma^{el}_{ij}=C_{ijkl} \Upsilon_{kl}$, where $C_{ijkl}$ is the elastic tensor and $\Upsilon_{kl}\equiv 1/2 \left(\nabla_i u_j+ \nabla_j u_i \right)$ the linear strain tensor. In all real solids (crystals with or without defects, glasses), there is a dissipative component $\sigma_{ij}'$ to the stress tensor due to the viscous component of the material response, which is proportional to the deformation rate. This contribution is ultimately due to anharmonicity of the lattice, and to finite temperature effects. Its structure is given by $\sigma_{ij}'=\eta_{ijkl}\partial_t \Upsilon_{kl}$ \cite{Chaikin}, and can be derived by symmetry arguments (hydrodynamics), or using the Rayleigh dissipation function~\cite{landau2013fluid,landau1986theory}. Here, $\eta_{ijkl}$ represents the viscosity tensor. For isotropic systems, the elastic and viscosity tensors can be written solely in terms of four parameters: the shear modulus $G$, the bulk modulus $K$, the shear viscosity $\eta$ and the bulk viscosity $\zeta$ (see \cite{Chaikin} for details). In the following, we will restrict ourselves to this situation and we will also neglect possible odd responses \cite{fruchart2022odd}.
After standard manipulations, the elastodynamic equations with the viscous dissipative contribution can be written in the form of a forced damped harmonic oscillator. The transverse Green's function in the mixed $(k,t)$ representation, $\mathcal{G}_{T}(k,t-t')$, is readily found by replacing the external force with a $\delta$-function source:
\begin{equation}
[\partial_{t}^{2} + (\eta/\rho)\,k^2\,\partial_{t} + (G/\rho)\,k^{2}]\, \mathcal{G}_{T}(k,t-t') = \delta(t-t') \label{tode}
\end{equation}
and upon Fourier-transforming in time we get:
\begin{equation}
\mathcal{G}_{T}(k,\omega)=\frac{1}{-\omega^{2} +(G/\rho)\,k^{2} -i\,\omega\,(\eta/\rho)\,k^{2}}\label{cinque}
\end{equation}
and, \textit{mutatis mutandis}, an analogous expression for the longitudinal Green's function.

In general, we thus have the Green's functions for longitudinal ($L$) and transverse ($T$) modes in the following generic form:
\begin{equation}
\mathcal{G}_{L,T}(k,\omega)=\frac{1}{\Omega^2_{L,T}(k)-\omega^2-i\,\omega\,\Gamma_{L,T}(k)},\label{Gf}
\end{equation}
with the poles providing the following set of dispersion relations for transverse and longitudinal phonons, respectively:
\begin{align}
&\omega_{L,T}= v_{L,T}\,k-i\,D_{L,T}\,k^{2}\,,\label{disp}\\
& v_{T}^2=\frac{G}{\rho}\,,\quad v_{L}^2=\frac{K\,+\,\frac{2\,(d-1)}{d}\,G}{\rho}\,,\\
&D_T=\frac{\eta}{2\,\rho}\,,\quad D_L=\frac{1}{2\,\rho}\left[\zeta+\frac{2(d-1)}{d}\eta \right]\,.
\end{align}
These expressions are valid only for low frequency/wavevector, but the higher order terms can be systematically derived using a perturbative scheme known as the \textit{gradient expansion}.
In general, $v_{L}>v_{T}$ since $K,G>0$ for all materials. Using Eq.\eqref{Gf} and Eq.\eqref{disp} we therefore identify $\Omega_{L,T}(k)=v_{L,T}\,k$ and
\begin{equation}
\Gamma_{L,T}(k)=2\,D_{L,T}\,k^2, 
\label{Akhiezer}
\end{equation}
\textit{i.e.}, a diffusive viscous damping, known as \emph{Akhiezer sound damping}~\cite{Akhiezer}. The root cause of Akhiezer damping is anharmonicity, as will be discussed in the following of this section.

\color{black}
Importantly, this framework recovers the ubiquitously observed finite temperature $\Gamma \sim k^{2}$ scaling of the acoustic phonon linewidth, which does not depend on the microscopic details of the system. As a matter of fact, Akhiezer damping perfectly reproduces the experimental data at low wavevector (see for example \cite{liao2020akhiezer} for a demonstration in a-Si$_3$N$_4$ and a-SiO$_2$ using the data of \cite{PhysRevMaterials.3.065601,Baldi2010,PhysRevB.77.100201,PhysRevB.71.172201,PhysRevB.14.823}).

The above derivation follows a hydrodynamic approach \cite{landau2013fluid} which is agnostic of the short-scale physics; by comparing with the result of a microscopic approach based on the Boltzmann transport equation for phonons, it has been shown that~\cite{Maris} 
\begin{equation}\label{DiffusionConstant-2}
D_{L} = \frac{C_{v} T \tau_U}{2\rho} \left(\frac{4}{3}\langle \gamma_{xy}^{2}\rangle -\langle \gamma_{xy} \rangle^{2} \right) \approx \frac{C_{v} T \tau_U}{2\rho} \langle \gamma_{xy}^{2}\rangle\,,
\end{equation}
where the last approximation for acoustic modes can be motivated with the typical wild fluctuations of $\gamma$ for low frequency vibrational excitations in both crystals \cite{Bongiorno} and metal alloys \cite{ZacconeWang}, hence $\langle \gamma_{xy} \rangle \approx 0$.
\\
Here, we neglected the contribution from the bulk viscosity $\zeta$, since normally $\eta \gg \zeta$. Furthermore,
$\langle...\rangle$ indicates averaging with respect to the Bose-Einstein distribution as a weight, while 
$\gamma_{xy}$ is the $xy$ component of the tensor of Gr{\"u}neisen constants. Also, $C_{v}$ is the specific heat at constant volume, while $\tau_U$ is the average time interval between two Umklapp scattering events. Since $\tau_U \sim T^{-1}$ (which is an experimental observation for most solids~\cite{Boemmel,Maris}), the diffusive constant $D_L$, and also the sound damping, are weakly dependent on temperature, \textit{i.e.}, a well-known experimental fact~\cite{Boemmel}, in the Akhiezer regime. 

A substantially equivalent expression for the damping of longitudinal phonons, in terms of an average Gr{\"u}neisen constant of the material $\gamma_{av}$, was proposed by Boemmel and Dransfeld~\cite{Boemmel},
\begin{equation}\label{DiffusionConstant-3}
D_{L} \approx \frac{C_{v} T \tau_U}{2\rho} \gamma_{av}^{2},
\end{equation}
and provides a good description of the Akhiezer damping measured experimentally in quartz at $T > 60K$ \cite{Boemmel}. 

In turn, the Gr{\"u}neisen constant $\gamma$, or at least the leading term~\cite{Krivtsov2011} of $\gamma_{av}$ or $\gamma_{xy}$ above, can be directly related to the anharmonicity of the interatomic potential. For perfect crystals with pairwise nearest-neighbour interactions, the following relation holds \cite{Krivtsov2011}
\begin{equation}
\gamma= -\frac{1}{6}\frac{V'''(a)a^{2} +2 [V''(a)a -V'(a)]}{V''(a) a+2 V'(a)},
\label{AtomicPotential}
\end{equation}
where $a$ is the equilibrium lattice spacing between nearest-neighbours, and $V'''(a)$ denotes the third derivative of the interatomic potential $V(r)$ evaluated at $r=a$. 
Hence, the phonon damping coefficient $D_{L}$ can be directly related to the anharmonicity of the interatomic potential via the Gr{\"u}neisen coefficient and Eq.\eqref{AtomicPotential}.\\

\subsubsection{Klemens damping of optical phonons}
The discussion in this subsection closely follows Ref.~\cite{Casella}.
On general grounds, the lifetime of a optical phonon can be rationalized by looking at its microscopic decay processes which are ultimately related to anharmonic (phonon-phonon) interactions. As proved in the pioneering work by Klemens \cite{Klemens}, the leading decay channel is the three-phonon scattering between two acoustic modes and an optical one, which can be described using Boltzmann formalism and many-body perturbation theory. Above room temperature, higher order processes become relevant as well and cannot be neglected anymore~\cite{Balkanski}. Despite the various approximations, Klemens result is in good agreement with controlled experimental observations~\cite{Kitajima}.

According to Klemens' theory, the mean lifetime of an optical phonon is given by
\begin{equation}
    \label{klemenstau}
        \dfrac{1}{\tau}=\omega\dfrac{J}{24\pi}\gamma^{2}_{G}\dfrac{\hbar\omega}{M v^{2}}\dfrac{a^{3}\omega^{3}}{v^{3}}C(\alpha,\beta)\left[1+\dfrac{2}{e^{x}-1}\right] 
\end{equation}
where
\begin{equation}
C(\alpha,\beta)=\dfrac{2}{\sqrt{3}}\dfrac{\alpha-\beta}{\alpha+\beta}\,;\,\,\,x=\dfrac{\hbar\omega}{2k_{B}T}\,.
\end{equation}
In the above formulae, $\omega$ refers to the phonon frequency in the undamped limit, and this applies to either longitudinal (LO) or transverse (TO) optical phonons. Furthermore,  $a^{3}$ is the volume per atom, $M$ is the ion mass, and $v$ is the acoustic phonon velocity within Debye approximation (since the decay process of the optical phonon may involve acoustic phonons). Also, $J$ is the label of the allowed channels by which the optical mode decays into  acoustic phonons,  and $\gamma_{G}$ is the Gr\"uneisen parameter of the solid related to anharmonicity of the interatomic potential.
Finally, $C$ is a coefficient on the order of $0.1$, which in the case of ionic crystals (e.g. alkali halides) depends on the spring constants of the two different ion species. This is because Klemens' original derivation focused on ionic crystals.

All these prefactors which appear in the Klemens formula Eq.\eqref{klemenstau} can be put into a single coefficient $\xi$, 
\begin{equation}
    \label{klemens_summarized}
    \dfrac{1}{\tau}=\omega^{5}\xi\left[1+\dfrac{2}{e^{x}-1}\right],
\end{equation}
where
\begin{equation}
\xi \equiv \dfrac{J}{24\pi}\gamma^{2}\dfrac{\hbar}{M v^{2}}\dfrac{a^{3}}{v^{3}}C(\alpha,\beta)\,,
\end{equation}
and $\omega$ the frequency of the optical phonon.
According to Klemens \cite{Klemens}, the above expression Eq.\eqref{klemens_summarized} can be further simplified. Using the Debye model, and approximating the optical phonon frequency with the Debye frequency $\omega_D$, one gets
\begin{equation}
\frac{1}{\tau}\propto \,\omega_D^2.
\end{equation}
Here, in good approximation, $\tau$ can be regarded as independent of the wavevector $k$. In the next sections, we will simply assume that the optical mode is not too dispersing and its frequency $\omega_{\text{opt}}$ can be approximated, at least in the limit of small wave-vector, by a constant. All in all, we will indicate as \emph{Klemens damping} the assumption that $\tau^{-1}\sim \omega^5_{\text{opt}}$, where $\omega_{\text{opt}}\equiv \mathrm{Re}\,\omega_{\text{opt}}(k=0)$, and $\omega_{\text{opt}}(k)$ is the dispersion relation of the optical mode. The main difference with the Akhiezer damping for acoustic phonons is that the Klemens damping is, at least in first approximation, independent of the wave-vector $k$, under the approximation detailed above.

 \subsection{The nature of the bosonic mediator}
 The starting point of the phenomenological theoretical framework is the definition of the mediator $\varphi$ responsible for the pairing and for the superconducting instability. For simplicity, we will focus on phononic mediators. Despite most of the treatment will identify the latter with acoustic/optical phonons, we will keep the derivation as general as possible to account for alternative bosonic mediators such as spin waves/magnons.
 
 The fundamental object under scrutiny is the retarded Green's function for the mediator $\varphi$ which in Fourier space takes the general form
 \begin{equation}\label{green}
 \mathcal{G}_\varphi(\omega,k)\,=\,\frac{1}{-\omega^2+\Omega^2(k)-i \omega \,\Gamma(k)}
 \end{equation}
 where $\omega,k$ are respectively frequency and wave-vector. Under few assumptions, this is the most general expression for the Green's function and $\Omega,\Gamma$ are, at this point, undetermined quantities which can be expanded in the low-energy limit, sometimes referred to as the \textit{hydrodynamic limit} or \textit{gradient expansion}, in a systematic power-series expansion in terms of $k$. Examples of this sort can be found in \cite{Kovtun:2012rj} for relativistic fluids, and in \cite{RevModPhys.95.011001} for phases of matter with broken translations. Importantly, this expansion is in general not convergent \cite{PhysRevLett.122.251601}.
 
 To continue, isotropy is assumed, where $k \equiv |\Vec{k}|$. The extension to anisotropic systems does not present any conceptual difficulties but it is rather cumbersome and therefore not explicitly shown. The poles of the retarded Green's function in Eq.\eqref{green} define the dispersion relation $\omega(k)$ of the corresponding excitations which is given by solving the following equation:
 \begin{equation}
     -\omega^2+\Omega^2(k)-i \omega \,\Gamma(k)\,=\,0\,.
 \end{equation}
 Here, we take the frequency $\omega$ to be complex and the wave-vector $k$ to be real.
 
 The Green's function presented in Eq.\eqref{green} corresponds to the following spectral function $\mathfrak{s}(\omega,k)$
 \begin{align}\label{spectral}
     \mathfrak{s}(\omega,k)&\equiv -\frac{1}{\pi}\,\mathrm{Im}\,\mathcal{G}_\varphi(\omega+i \delta,k) \nonumber\\ &=\frac{\omega\,\Gamma(k)}{\pi\,\left[\left(\omega^2-\Omega^2(k)\right)^2+ \omega^2 \Gamma^2(k)\right]}\,,
 \end{align}
 which is directly accessible via scattering experiments and shows the typical Lorentzian shape. 
 
 In the following, we will consider two fundamentally different types of excitation. First, we discuss the scenario in which the mediator corresponds to a gapless mode whose dispersion relation at low-energy is given by
 \begin{equation}\label{dispacu}
 \Omega(k)=v \,k+\dots\,,\qquad \Gamma(k)=D\, k^2+\dots
 \end{equation}
 where the $\dots$ indicate higher-order corrections in the wave-vector $k$. Here, $v$ defines the propagation speed while $\Gamma= D k^2$ the diffusive sound attenuation. By abusing the language, we will refer to $D$ as the diffusion constant. Intuitively, Eq.\eqref{dispacu} can be identified as the low-energy solution of a dynamical equation of the type:
 \begin{equation}\label{visceq}
     \frac{\partial^2 \phi}{\partial t^2}+ v^2 \frac{\partial \phi^2}{\partial x^2} + D\, \frac{\partial}{\partial t}\, \frac{\partial \phi^2}{\partial x^2}=0
 \end{equation}
 with $ \phi(t,x)=e^{-i \omega t+ i k x} \phi_0$, where using isotropy we have assumed the spatial dependence to be only along the $x$ direction. Eq.\eqref{visceq} must be taken with a grain of salt since low-energy sound modes usually appear in the context of hydrodynamics from a more complicated dynamics in terms of coupled fluctuations, \textit{e.g.}, particle number, momentum, energy, etc and not from the dynamics of a single low-energy variable.
 
 The collective variable $\phi$ is the dynamical field corresponding to the mediator $\varphi$.
 The most notable example obeying a dispersion as in Eq.\eqref{dispacu} is that of acoustic phonons. In this concrete case, Eq.\eqref{visceq} corresponds to the dynamical equation obtained from viscoelasticity theory where $\phi$ is identified with the infinitesimal displacement field \cite{Chaikin}. For transverse (T) and longitudinal (L) acoustic phonons, one obtains (\textit{cfr.}, Section \ref{lolo} above for the derivation):
 \begin{align}
& v_{T}^2=\frac{G}{\rho}\,,\quad v_{L}^2=\frac{K\,+\,\frac{2\,(d-1)}{d}\,G}{\rho}\,,\\
&D_T=\frac{\eta}{\rho}\,,\quad D_L=\frac{1}{\rho}\left[\zeta+\frac{2(d-1)}{d}\eta \right]\,,
\end{align}
where $G,K,\eta,\zeta,\rho$ are respectively the static shear modulus, the static bulk modulus, the shear viscosity, the bulk viscosity and the mass density of the system \cite{Chaikin}. We have defined with $d$ the number of spatial dimensions and neglected the effects from thermal expansion. For the moment, we will be agnostic about the microscopic origin of the damping term $\Gamma$ and we will take $v,D$ as pure phenomenological parameters. Because of stability requirements, we have $v^2,D>0$. 

\begin{figure}
    \centering
    \includegraphics[width=0.45\linewidth]{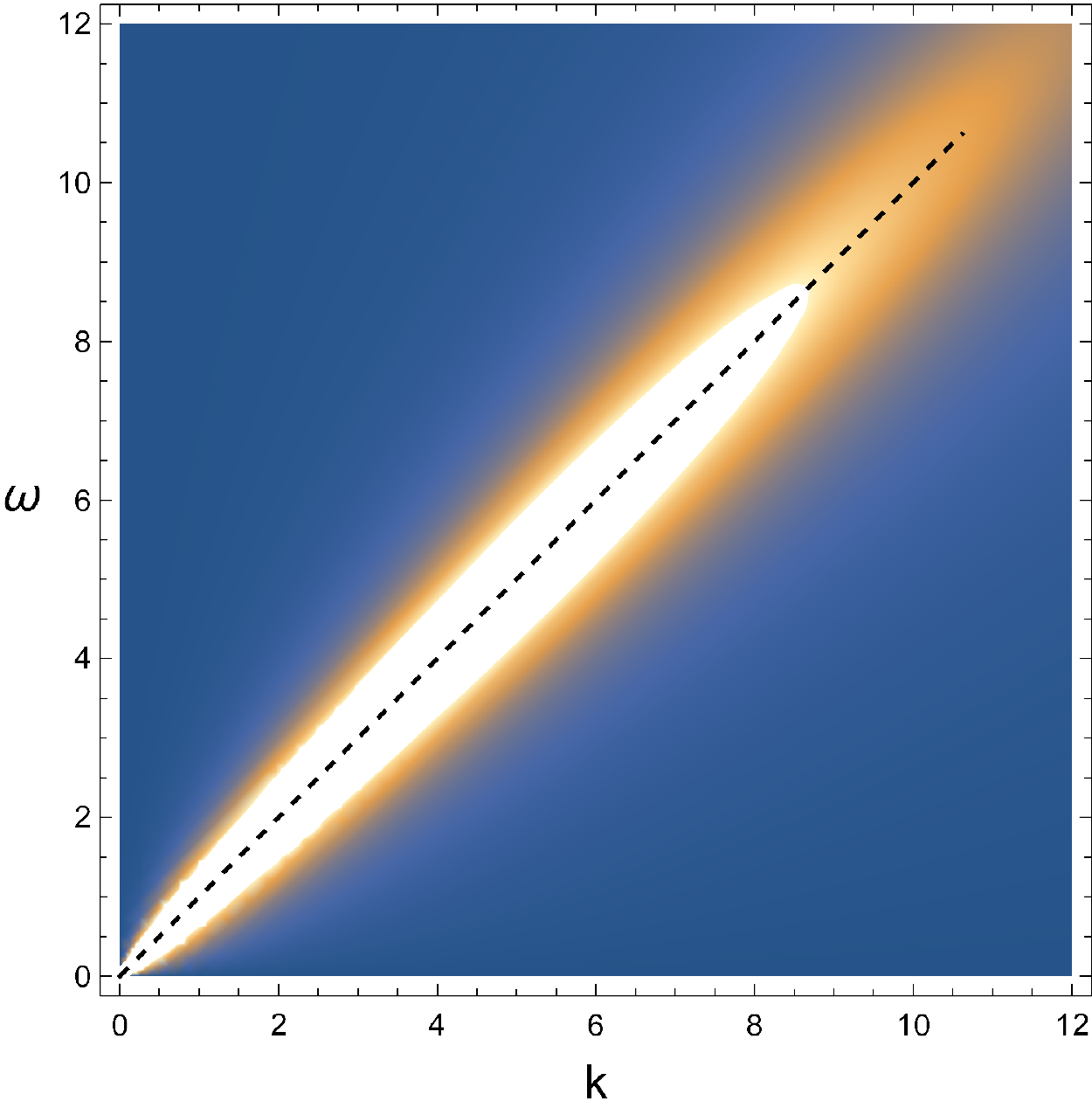} \qquad 
    \includegraphics[width=0.45\linewidth]{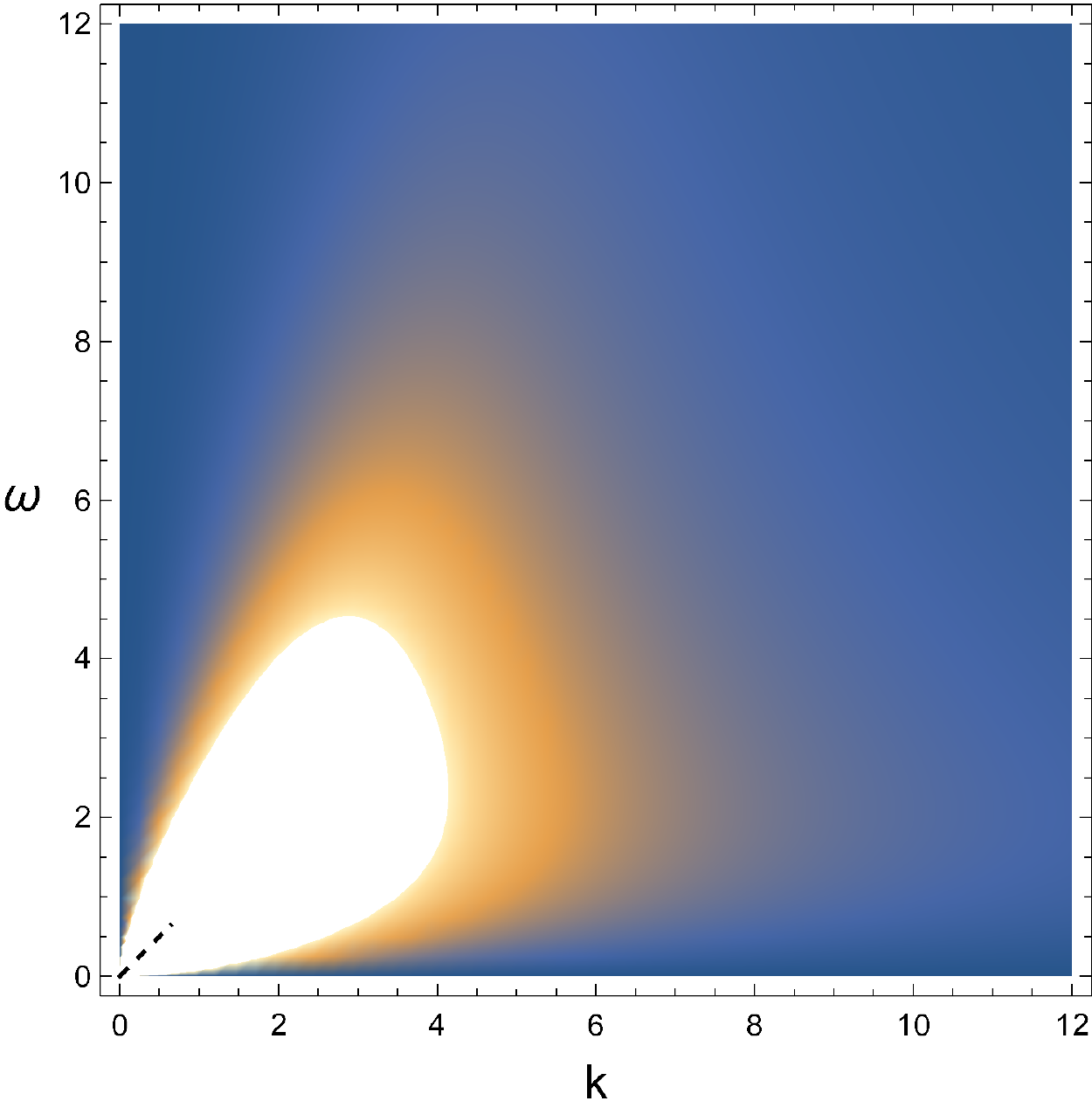}
    \caption{The spectral function for an underdamped acoustic mode using Eq.\eqref{spectral} and the parameters defined in Eq.\eqref{dispacu}. The dashed line shows the real part of the dispersion relation $\omega=v k$ up to the Ioffe-Regel scale $k^\star$. The speed of sound is taken to unity $v=1$ while the diffusion constant $D=0.03,1$ (left, right).}
    \label{fig:acu}
\end{figure}

In this first situation, a fundamental scale in the problem is given by the so-called Ioffe-Regel (IR) wave-vector $k^\star$ \cite{ioffe1960non}, defined as the root of:
\begin{equation}
    \Omega(k^\star)=\pi\, \Gamma(k^\star)\,.
\end{equation}
The IR scale qualitatively indicates the energy at which the acoustic mediator loses its well-defined propagating nature and turns into a diffusive quasi-localized mode. Physically, the larger the attenuation constant $\sim D$, the lower the energy at which the coherent nature of the mediator is lost. This is evident in Fig.\ref{fig:acu} in which the spectral function of the bosonic mediator is shown for two very different values of $D$.

A second relevant scenario is defined by the following alternative choice
 \begin{equation}\label{opt}
     \Omega(k)=\omega_0+ \alpha\, k^2+\dots\,,\qquad \Gamma(k)=\Gamma_0+\dots
 \end{equation}
 where $\omega_0$ represents the energy gap (the "mass" in particle physics jargon) and $\Gamma_0$ the wave-vector independent scattering rate. The parameter $\alpha$ takes into account the eventual mild $k$ dependence in the dispersion relation of the mode. Finite values for $\omega_0,\Gamma_0$ are prohibited for acoustic phonons because of their Goldstone mode nature but they can naturally appear once one considers optical modes which are not protected by any fundamental symmetry breaking pattern. While the sign of $\Gamma_0$ is fixed by stability arguments to be positive, the one of $\alpha$ is a priori undetermined and strongly dependent on the microscopics of the system.
 
 A simplified possibility is to neglect the $k$ dependence in Eq.\eqref{opt}, and consider a simpler dispersion relation:
 \begin{equation}\label{optdisp}
     \omega^2=\omega_0^2-i \,\omega \,\Gamma_0\,.
 \end{equation}
 Once more, depending whether $\mathrm{Re}(\omega)>\mathrm{Im}(\omega)$ or viceversa, the dynamics will result underdamped or overdamped. The transition roughly happens when $\omega_0 \sim \Gamma_0$. Therefore, we find convenient to define a dimensionless parameter:
 \begin{equation}
     \Tilde{\Gamma}\equiv \frac{\Gamma_0}{\omega_0},
 \end{equation}
 such that in the regime $\Tilde{\Gamma}\ll 1$ we have a coherent well-defined bosonic quasiparticle mediating the pairing, while for $\Tilde{\Gamma}\gg 1$ the mediator becomes incoherent and does not correspond anymore to a well-defined quasiparticle.\\
 Notice that in general, effective parameters such as $\omega_0,\Gamma_0$ are implicit functions of thermodynamic variables (temperature, doping, etc.). In order to reveal their explicit dependence a microscopic theory is needed. Nevertheless, we will see that in some scenarios (for example the soft mode instability mechanism described in Section \ref{secsoft}) one can simply introduce a convenient parameterization and obtain interesting physical results.\\
 
Later in the paper, we will use the results of this section to examine the manner in which superconducting properties, like critical temperature, are affected by the damping and other parameters appearing in the dispersion relations, Eqs.\eqref{dispacu}-\eqref{opt}.
 \section{Damped Bosons: Experimental Probes}
 \begin{figure}
    \centering
    \includegraphics[width=0.8\linewidth]{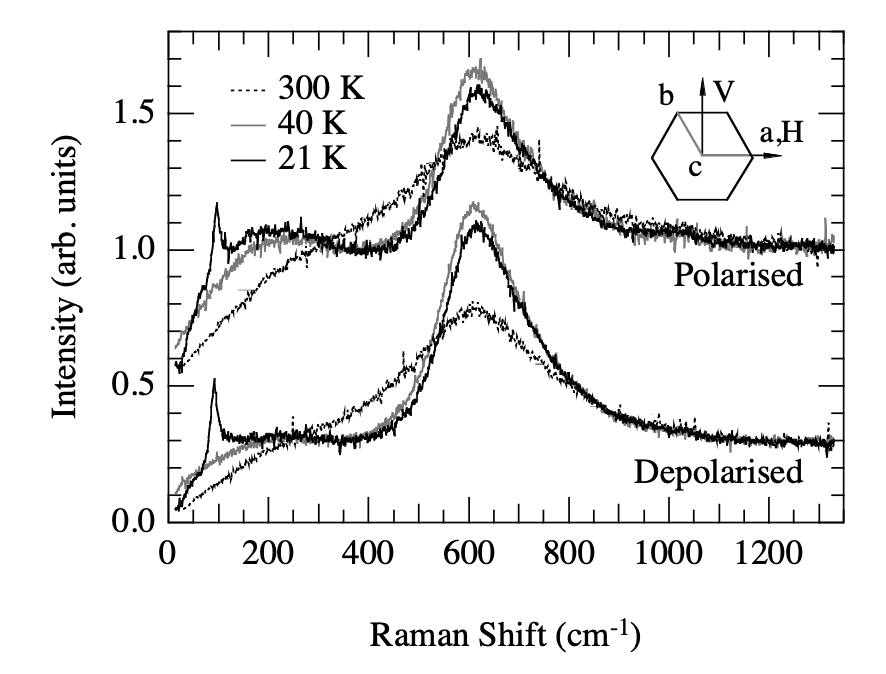}
    \caption{Raman scattering spectra showing the broad E$_{2g}$ mode at $\sim 620$cm$^{-1}$  in MgB$_2$ attributed to anharmonic phonon decay. Copyright APS from Ref.~\cite{quilty2002superconducting}. }
    \label{fig:Raman}
\end{figure}
 In this section we review various experimental probes used to quantify anharmonic damping in the pairing mediators. This can be achieved either by fitting the spectral line shape of the boson or by directly measuring their correlation functions. Typically, these observables are measured as a function of an external control parameter like pressure, carrier concentration, impurities, temperature etc. We broadly classify the probes according to their coupling to the charge, as in lattice based mediators like phonons, or to the spin, as in spin based mediators such as spin fluctuations. The list of experiments discussed in each category below is by no means comprehensive but rather a collection of representative examples that allows the interested reader to further explore each topic. We begin with probes of anharmonic damping in phonons. 
 \subsection{Raman scattering}
 Raman scattering is the most widely used technique to measure phonon properties (frequency shifts and linewidths) in quantum materials~\cite{menendez1984temperature, devereaux2007inelastic}. The method involves measuring the frequency shift of visible light scattered from a sample at zero momentum. The anharmonic damping is then extracted by fitting the line shape with well known anharmonic models that contribute to the linewidth~\cite{cowley1964theory, cowley1965anharmonic, cowley1968anharmonic, menendez1984temperature}.Additionally, the polarization of light can be varied to access different symmetry channels of the crystal point group of the specific materials. In principle, several scattering processes contribute to the phonon linewidth. These include electron-phonon, multi-phonon, impurity scattering, lattice dislocations etc.,  and care must be taken to extract the purely anharmonic component. The classic BCS superconductor MgB$_2$ serves as an illustrative example of extracting phonon anharmonicity using Raman scattering (see Fig.~\ref{fig:Raman}). Here, the E$_{2g}$ phonon mode, centered around $\sim 620$cm$^{-1}$, has a large anomalous broadening of $\sim 200-280$cm$^{-1}$~\cite{blumberg2007multi,goncharov2001raman, quilty2002superconducting, mialitsin2007anharmonicity, renker2003strong}.  The experiments were performed on clean samples and the electron-phonon coupling was calculated to account for only about $50$cm$^{-1}$. The remaining scattering was attributed to multi-phonon decay from anharmonic effects (see~\cite{blumberg2007multi} and references therein). More recently, similar broadening effects and anharmonic frequency shifts were also noticed in high pressure superconductors such as hydrides~\cite{Eremets_review,Eremets2015,Dias2021-PRL} and TlInTe$_2$~\cite{yesudhas2020origin}, and non-stoichemotric Fe based chalcogenide  superconductors K$_y$Fe$_{2-x}$(Se,S)$_2$~\cite{ignatov2012structural, lazarevic2011phonon}. 
 We will further discuss the relationship between superconductivity and anharmonicity for the specific case of TlInTe$_2$ later in this review. 
 \begin{figure}
    \centering
    \includegraphics[width=0.8\linewidth]{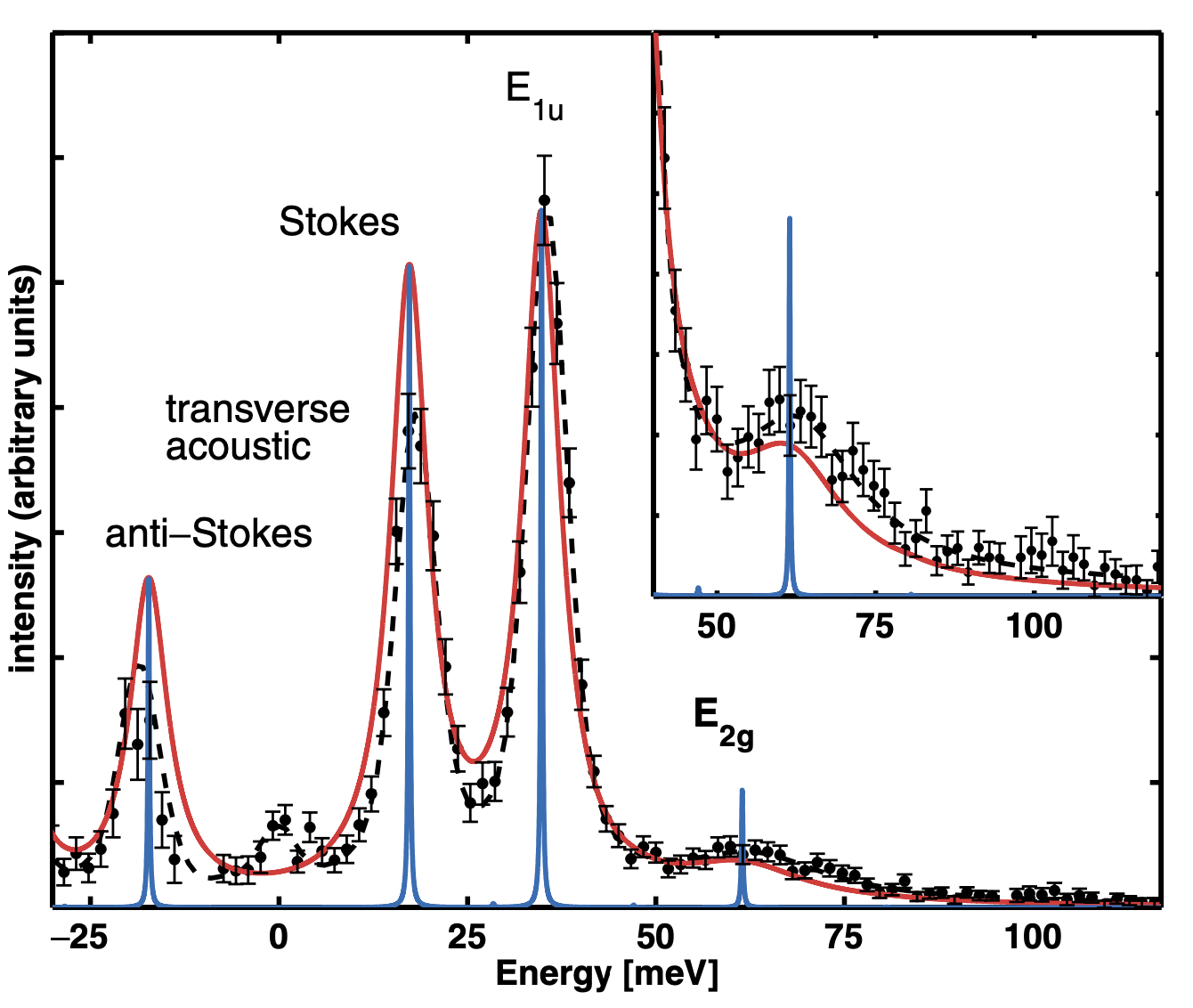}
     \includegraphics[width=0.8\linewidth]{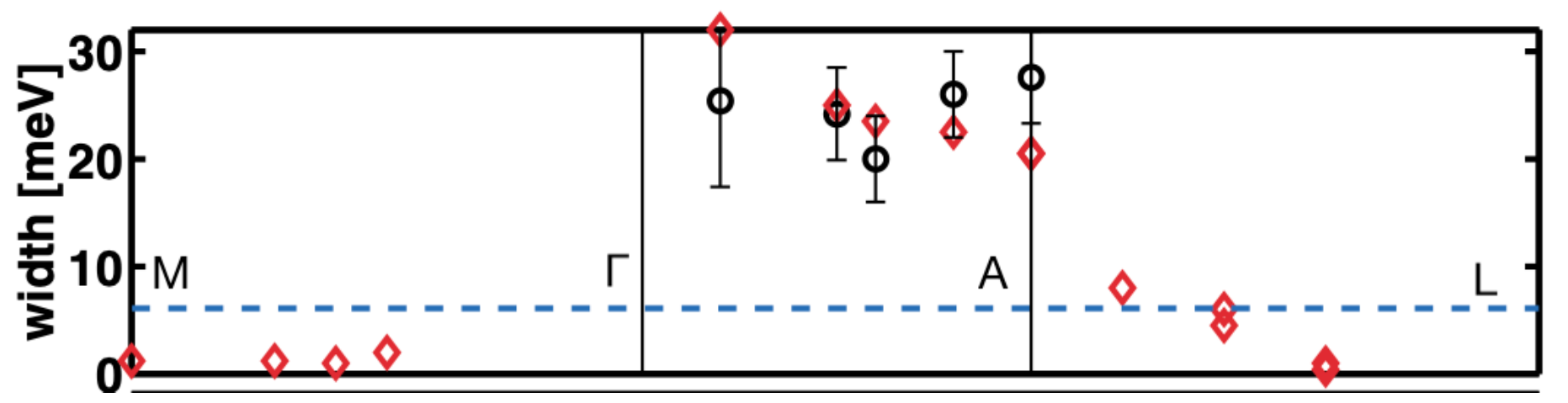}
    \caption{IXS spectra in MgB$_2$. (Top) Symmetry decomposed data with the E$_{2g}$ mode centered around 60 meV taken at a momentum point 0.6 $\Gamma-A$. (Bottom) Momentum resolved width of the E$_{2g}$ mode enhanced in the $\Gamma-A$ direction. Copyright APS from Ref.~\cite{shukla2003phonon}. }
    \label{fig:IXS}
\end{figure}
 \subsection{Inelastic x-ray scattering}
 Another widely used tool to extract anharmonic damping effects in phonons involves inelastic X-ray scattering (IXS) (see Refs.~\cite{burkel2000phonon,krisch2006inelastic, baron2009phonons} for a review on phonon spectroscopy using IXS). The IXS technique is also a photon-in/photon-out process like Raman scattering, but is performed at higher (X-ray) energies and is momentum resolved. Here an incoming photon with a given energy and momentum scatters off a phonon in the sample to create an outgoing photon with a different energy and momentum. The frequency shift and linewidths at different momentum transfers are then fit to theoretical models to extract anharmonic damping effects. Like Raman scattering, polarization of the incoming photons can be manipulated to access different symmetry channels of the solid. In this regard, IXS has become an invaluable probe for mapping out phonon dispersion relations and extracting momentum and symmetry dependent spectral lineshapes~\cite{burkel2000phonon,krisch2006inelastic, baron2009phonons}. IXS spectra for various metals and superconductors have been summarized in~\cite{burkel2000phonon, krisch2006inelastic, baron2009phonons}. Returning to the prototypical example of MgB$_2$, IXS spectra and phonon dispersions and linewidths were studied in Refs.~\cite{shukla2003phonon,baron2004kohn, d2007weak}. Momentum resolved data suggested that the broad linewidth of the $E_{2g}$ phonon mode was dominated by the $\Gamma$-A direction of the Brillouin zone (see Fig.~\ref{fig:IXS}).  The contribution to the linewidth from phonon anharmonicity was also found to be smaller than the electron-phonon broadening in apparent contradiction of earlier Raman scattering studies~\cite{quilty2002superconducting}. This disagreement between IXS and Raman scattering was later addressed in Ref.~\cite{d2007weak}. More recently, it was demonstrated that the E$_{2g}$ mode is strongly coupled to electrons and higher-order electron-phonon scatterings become relevant leading to large effective phonon-phonon dampings at zero momentum~\cite{Dovko2018,Dovko2020}. As a result, anharmonic linewidths are sometimes difficult to separate from the usual electron-phonon contribution, since phonon-phonon scatterings mediated by higher-order electron-phonon scatterings have similar temperature dependence~\cite{Dovko2018,Dovko2020}. Finally, phonon linewidths from IXS with momentum resolution have also been measured in other superconducting systems such as cuprates~\cite{uchiyama2004softening, fukuda2005doping, graf2008bond, le2014inelastic}, iron superconductors~\cite{fukuda2008lattice}, soft phonon systems like CaAlSi~\cite{Akimitsu2008} 
 as well as the newly discovered kagome superconductors~\cite{li2021observation}. 
 \subsection{Electron energy loss spectroscopy (EELS)}
Our focus so far was on purely photonic probes. We now turn to electronic scattering methods to probe phonon damping on material surfaces and thin films. Electron energy loss spectroscopy (EELS) is a popular technique in this category where electrons with particular energy (and momentum, as is the case in the  momentum resolved counterpart M-EELS~\cite{vig2017measurement}) are shot into the sample to determine the nature of surface phonons~\cite{ibach2013electron}. The energy (and momentum) transfer to the sample is then determined from the kinematics of the scattered electron. In superconductors, EELS has been used predominantly to extract medium to high energy ($\sim 30$meV$-1$eV) electronic properties and response functions. In principle, information of phononic lineshapes can also be obtained depending on the energy resolution of the device; although, to our knowledge, a systematic study that isolates the effects of mode specific anharmonic phonon damping on the lineshape is currently missing. This is due to the fact that coupling of electronic probes to specific phonon symmetry modes is not straightforward with high resolution unlike photon based probes. Early EELS data in cuprates~\cite{phelps1994absence} studied and modeled  surface optical phonon line shapes.  Recently, M-EELS measurements in the normal state of the Cuprates~\cite{mitrano2018anomalous} were analysed~\cite{huang2021extracting,setty2018inequivalence}  to isolate the phononic and electronic components, and determine their independent and combined effects on  correlation properties of the strange metal. In MgB$_2$~\cite{sahadev2012high}, several phonon excitation energies were associated to features obtained in the EELS spectra.  In strontium ruthenate~\cite{wang2017quasiparticle}, the bulk and surface phonon lineshapes, and their coupling to quasi-1D electronic bands was explored.   However,  neither phonon broadening due to anharmonic damping nor its relationship to superconductivity were systematically studied in these works. \par
\begin{figure}
    \centering
    \includegraphics[width=0.8\linewidth]{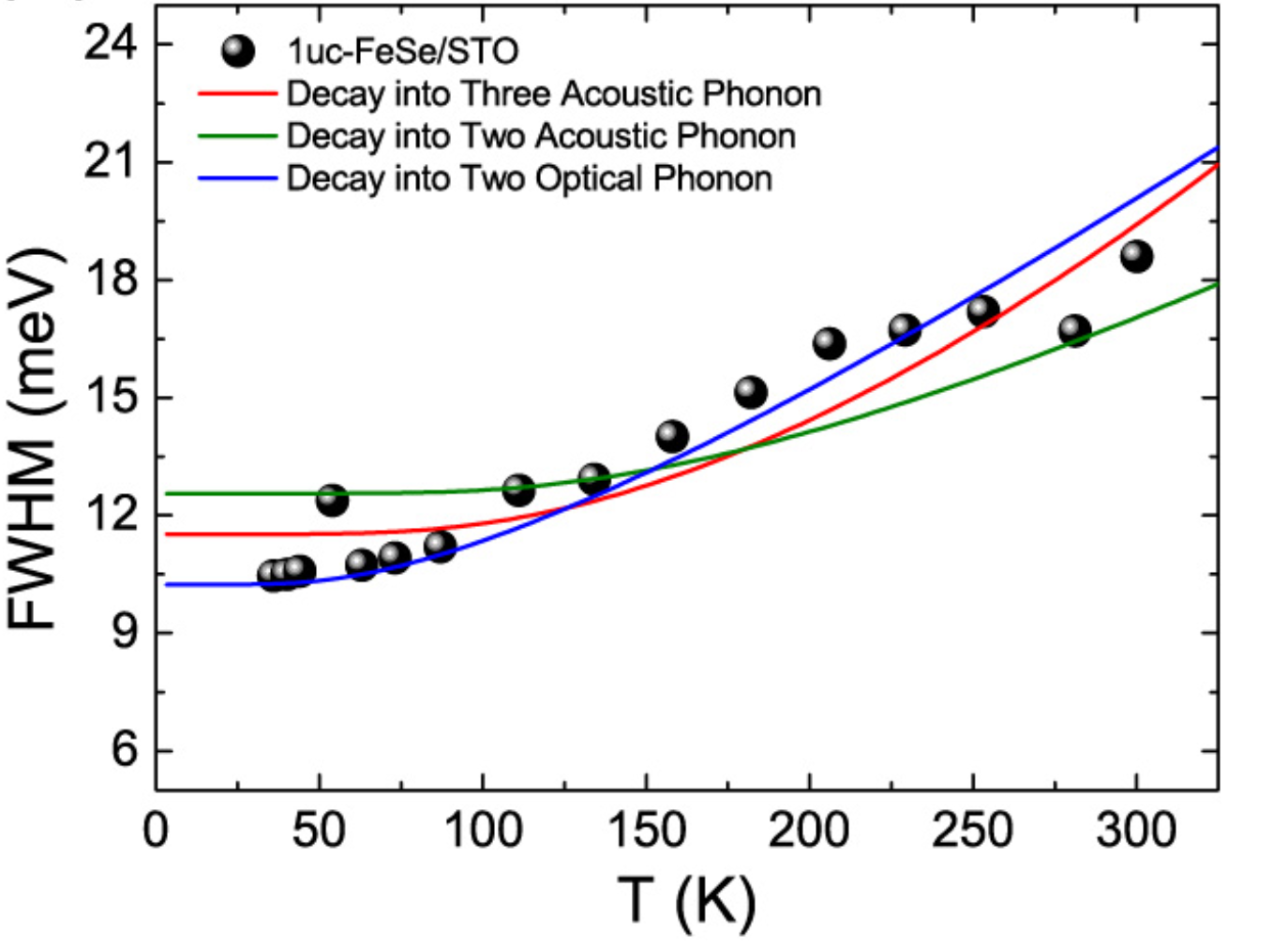}
    
    \vspace{0.1cm}
    
     \includegraphics[width=0.8\linewidth]{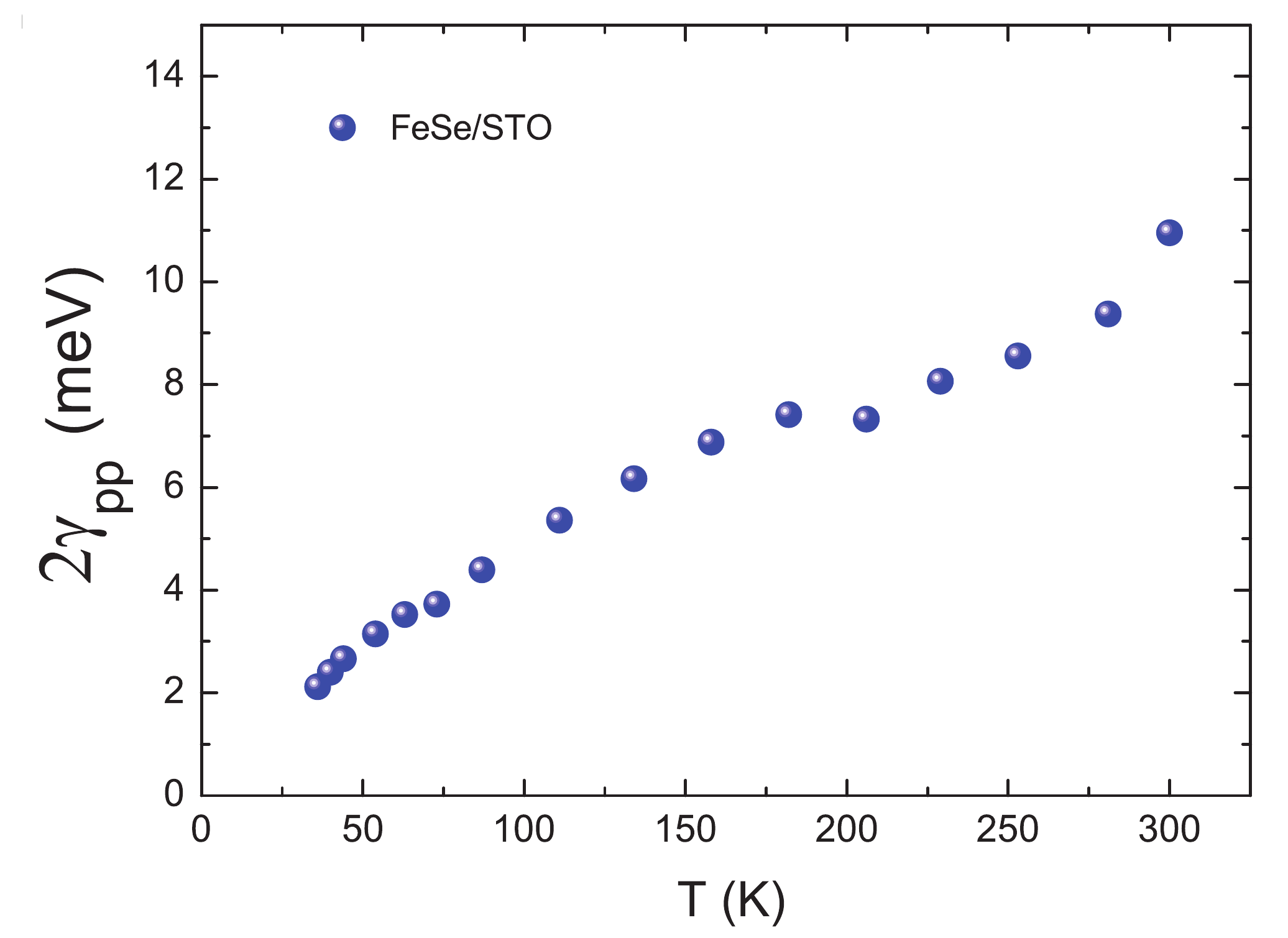}
    \caption{\textbf{(Top)} Fits of the full width at half maximum (FWHM) for different anharmonic models with multi-phonon decay channels in single unit cell FeSe on strontium titanate substrate. \textbf{(Bottom)} Extracted phonon-phonon contribution to the FWHM by subtracting the $T=0$ electron-phonon component in the two optical phonon anharmonic model. Copyright APS from Ref.~\cite{Guo2016}. 
    }
    \label{fig:EELS}
\end{figure}

 A comprehensive study of phonon broadening due to anharmonic decay in single unit cell FeSe films on strontium titanate was examined by the authors of Ref.~\cite{Guo2016}. Properties of specific phonon frequency branches ($\alpha$ and $\beta$ modes) such as the  energy and full width at half maximum (FWHM) as a function of temperature was modelled.  The  anharmonic contribution to the FWHM from multi-phonon decay processes was obtained by subtracting the $T=0$ (electron-phonon) contribution.  Fig.~\ref{fig:EELS} shows the plots of the total FWHM (top panel) and the extracted anharmonic component (bottom panel).  
 \subsection{Probes of Gr\"{u}neisen constant}\label{opop}
 
The importance of the Gr\"{u}neisen constant or Gr\"{u}neisen parameter as a quantitative estimate of the extent of anharmonicity in superconductors has been recently emphasized. Since the Gr\"{u}neisen constant essentially describes how the acoustic phonon frequencies change with volume, it can be measured by mechanical ways by linking to the nonlinear elastic behaviour of the solid.
 
 Gilvarry \cite{gilvarry1957,PhysRev.102.331,gilvarry1956equation} was able to connect the Murnaghan equation of state of nonlinear elasticity (linking changes in pressure to changes in volume) to the traditional Gr\"{u}neisen assumption that the normal mode frequencies $\omega_i$  of the lattice model for particles exhibiting anharmonic interactions should have the volume dependence,
\begin{equation}
    \gamma_i=\frac{\partial \ln \omega_i}{\partial V} \label{e2}
\end{equation}
 where $V$ is the material volume so that $\gamma_i$ describes how the normal mode frequencies change with material volume, regardless of the detailed molecular origin of the volume change.  For reference, the Gr\"{u}neisen exponent $\gamma_i$ for an ideal harmonic (Debye) lattice material equals $1/3$, \textit{i.e.}, for all of the normal modes. More generally, the Gr\"{u}neisen parameter $\gamma_G$, represents an average over the normal modes of the material, so that $\gamma_G$ normally differs from 1/3 in materials having more realistic intermolecular interactions. Gilvarry's anharmonic extension of the Debye model assuming Eq.~\eqref{e2} and a constancy of the Poisson ratio, leads exactly to Murnaghan's equation of state where the scaling exponent $\gamma_M$ equals,
\begin{equation}
    \gamma_M= 2\, \gamma_G +\frac{1}{3}\,,
\end{equation}   
which provides a link between the microscopic atomic dynamics and macroscopic elasticity. A power-law scaling of the normal mode frequencies with $V$ was originally motivated by schematic choices of anharmonic interparticle potentials (Mie or Lennard-Jones) where the repulsive and attractive contributions to this potential have variable power exponents. This was already considered by Gr\"{u}neisen \cite{gruneisen1912theorie} and many others \cite{gilvarry1957,PhysRev.102.331,gilvarry1956equation}. In general, the Gr\"{u}neisen  parameter $\gamma_G$ can be specialized to particular normal modes \cite{PhysRevLett.124.215501} or mode types (longitudinal or transverse) ~\cite{Wang2003}. Often, however, an appropriately defined average over all the modes of the system is assumed. This approximation seems to be particularly suitable for glass-forming liquids and amorphous solids where the existence of well-defined normal modes of the type found in crystals is not so well defined. 

We then have a semi-empirical equation of state generalizing the Debye theory in which there is an explicit link between the microscopic measure of anharmonicity $\gamma_G$  and the macroscopic measure derived empirically from nonlinear elasticity.  The importance of the Murnaghan equation in understanding the temperature dependence of relaxation in condensed materials has become appreciated \cite{doi:10.1021/acs.macromol.1c00075}.
From a thermodynamic perspective, $\gamma_G$ describes the rate of change of the pressure as the internal energy varies at a constant volume and this interpretation leads to an explicit expression in terms of the specific heat $C_V$, the thermal expansion coefficient and the isothermal compressibility \cite{gilvarry1957,PhysRev.102.331,gilvarry1956equation,gilvarry1955},

There are extensive tabulations of $\gamma_G$ measured experimentally  and the many properties to which it is interrelated \cite{RevModPhys.38.669,anderson1966,barker1967,sharma1983relationship}. The application of $\gamma_G$ in materials science has been discussed recently in  \cite{anderson2000,de2016thermodynamics}. 
Recently, there have been significant advances in the first principle computation of the average $\gamma_G$  and the Gr\"{u}neisen exponent for particular modes \cite{PhysRevLett.124.215501,yang2021giant}.

 \subsection{Inelastic Neutron Scattering (INS)}
Neutrons scattering off lattice vibrations forms another complementary probe of phonon dispersions and linewidths (see Refs.~\cite{waller1952neutron, elliott1967group} for early conceptual work). Unlike the previous probes, neutrons do not couple through the charge due to their charge neutrality. Rather the coupling to the lattice occurs through atomic displacements via interactions with the nuclei. These interactions are  typically modeled with a short range ``hard core" isotropic potentials~\cite{hudson2006vibrational}. Like IXS and EELS, neutron scattering is capable of extracting mode and momentum resolved phonon dispersions and linewidths. Earlier INS work in Nb$_3$Sn by Axe and Shirane~\cite{axe1973inelastic} found abrupt changes in the lifetimes of certain transverse acoustic phonon modes near the superconducting transition temperature. They further discussed certain empirical relationships between superconductivity and damping induced by anharmonicity and electron-phonon coupling.  In liquid helium, Ref.~\cite{cowley1971inelastic} used INS to  extract phonon linewidths at various temperatures and wave vectors, and classified the total INS structure factor and dampings according to  one-phonon and multi-phonon contributions. In the cuprates, early INS studies~\cite{reichardt1994anharmonicity, chou1990inelastic} laid out the role of anharmonic phonon damping and electron-phonon coupling to the linewidths of various phonon modes as well as their relationship to superconductivity.  In the iron based superconductor CaFe$_2$As$_2$, large phonon linewidths were found~\cite{mittal2009measurement} from INS opening up a possible role for anharmonicity in the pnictides.  Ref.~\cite{yamaura2019quantum} assigned excitation of two distinct phonon modes to different types of extremely anharmonic phonons arising from ``quantum rattling" in deuterium doped LaFeAsO.  In the YNi$_2$B$_2$C superconductor, momentum resolved INS phonon linewidths were studied in Refs.~\cite{Hradil2008, weber2014phonons}. These authors concluded that the scattering rate was dominated by electron-phonon coupling rather than anharmonicity. Finally, in MgB$_2$, first-principles calculations of lattice dynamics were performed and found to be in agreement with INS data~\cite{yildirim2001giant}. In this work, a giant anharmonicity of the E$_{2g}$ in-plane boron phonons and nonlinear electron-phonon coupling was found to be important for understanding superconductivity.  Anomalous behavior of the phonon density of states due to multi-phonon processes in MgB$_2$ was further explored in Ref.~\cite{muranaka2002vibrational}.    
 \subsection{Atomic scattering}
 Scattering of inert gas atoms such as Helium over crystal surfaces is another tool to probe properties of surface phonons~\cite{cabrera1969theory, manson1971inelastic, Toennies1994}. The technique involves inert gas atoms with an initial momentum and energy incident on a crystal surface, and interacting with lattice vibrations via a generic two-body atomic potential.  The energy and momentum transferred to the phonons is measured from the inelastically scattered atoms; the kinematics and decay of the vibrations can then be mapped out using this information. Phonon dispersions obtained from surface scattering using He atoms for most part agree with EELS measurements, and the two techniques complement each other in covering much of the surface phonon vibrational spectra~\cite{Toennies1994}.  In superconductors, Helium atom scattering has been used over the last several years to extract the electron-phonon coupling constant~\cite{sklyadneva2011mode, benedek2014unveiling, benedek2020measuring, gloria2021electron}.  Ref.~\cite{sklyadneva2011mode}, for example, used inelastic Helium atom scattering  to measure electron-phonon coupling strengths for each phonon mode in superconducting Pb films. The mode/momentum specificity of the couplings has yet to be properly exploited to study other superconducting families.   Obtaining anharmonic effects including anharmonic phonon damping in superconductors using atomic scattering has been less explored. One rare example~\cite{gester1994combined} is the case of metallic Aluminum where Helium atom scattering was used to obtain surface-phonon anharmonicity and linewidths on the Al(100) and Al(111) surfaces. The results were shown to be in good agreement with molecular dynamics simulations over a range of wave vectors and temperatures. Over the last decade, due to the utility of atomic scattering to study properties surface phonons, the coupling between Dirac fermions and phonons on the  surface of the strong topological insulators has also been studied in a mode-specific manner~\cite{zhu2011interaction, Batanouny2012}. We anticipate applications of this technique  to measure low-lying `topological phonon'~\cite{stenull2016topological, liu2020topological} surface modes and their potential relationship to superconductivity in the near future~\cite{di2022theory}.   
 \subsection{Point contact spectroscopy}
  Over the last several decades, the capability of designing nanometer size orifices at the junction of metals and superconductors has enabled a new spectroscopic tool to probe electronic properties. Termed as point contact spectroscopy (PCS), such an experimental geometry has been successful in quantifying various electronic relaxation mechanisms in metals, superconductors, heavy fermion systems (see Refs.~\cite{Yanson2005, Wyder1980, Naidyuk-Yanson} for detailed reviews) and non-fermi liquids~\cite{Phillips-PNAS}. Relevant to our discussion is the role played by PCS in extracting energy resolved electron-phonon interactions in metals and superconductors. 
  For normal metals, the basic geometry consists of a nanometer(s)-thick dielectric layer that separates 
 two metallic films. The dielectric layer contains a small constriction with a diameter of the order of the scattering length of the electron injected into it. The resistance of the `point' contact is given approximately by the interpolation formula
  \beq
 R &\simeq& R_{sh} \left( 1+ \frac{3 \pi d}{16 v_f \tau_{e-ph}} \right) ,\\
 \tau_{e-ph}^{-1} &=& \frac{2 \pi }{\hbar} \int_0^{eV} d\omega~
\alpha^2(\omega) F(\omega).
\label{Eq:Resistance}
 \eeq
Here $R_{sh} = \frac{16 \rho_r l}{3\pi d^2}$ is the Sharvin resistance, $\rho_r$ is the resistivity, $l$ ($\tau_{e-ph}^{-1}$) is the electron-phonon scattering length (rate), $v_f$ the Fermi velocity, $d$ the diameter of the constriction and $eV$ the bias voltage. The information of electron-phonon coupling is contained in $\alpha(\omega)$ and the phonon density of states is given by $F(\omega)$. The Eq.~\ref{Eq:Resistance} is an interpolation formula for the contact resistance between the clean ($l>d$; dominated by $R_{sh}$) and dirty ($l<d$; dominated by Maxwell resistance $R_M \equiv \rho_r/d$) limits.

The derivative of the bias dependent contact resistance is proportional to the second derivative of the voltage with respect to the current and is given by 
 \beq
 \frac{dR}{dV} \propto \frac{d^2V}{dI^2} = R_{sh}\frac{3\pi^2 e d}{8\hbar v_f} \alpha^2F(eV).  
 \label{Eq:ResistanceDerivative}
 \eeq
The key quantity on the right hand side is the Eliashberg function $\alpha^2F(eV)$ which is the convolution of the phonon density of states and the electron phonon coupling function. Thus from the derivative of the contact resistance, information about the phonon linewidths can be obtained, and the method has been applied to a wide variety of metals to extract the Eliashberg function~\cite{Yanson2005}. A formal theory justifying Eq.~\ref{Eq:ResistanceDerivative} appears in Refs.~\cite{Yanson1983, Kulik1992}. Returning to our prototypical case of  MgB$_2$, a damped maximum was found above 60 meV (width $\sim 15 $meV) consistent with the E$_{2g}$ phonon modes~\cite{Naidyuk2003-MgB2,Naidyuk2004-MgB2,Jansen2003-MgB2} observed in other probes. To our knowledge, there is currently no systematic PCS study that separates the linewidth contributions originating from anharmonic and electron-phonon interaction effects. An effort in this direction could greatly complement existing Raman and neutron scattering analyses discussed in previous subsections. 
 \subsection{Spin based techniques}
\begin{figure}
    \centering
    \includegraphics[width=0.8\linewidth]{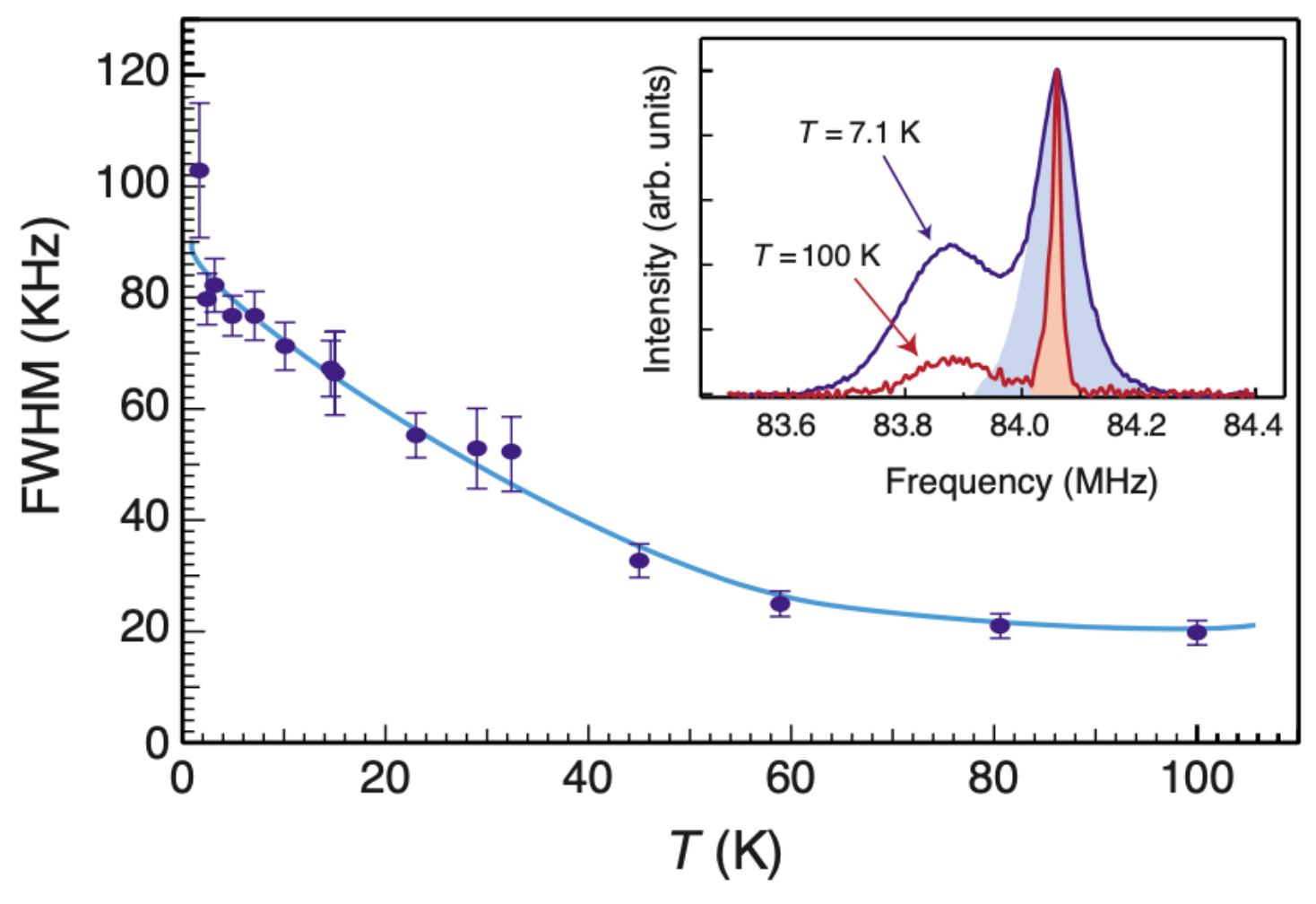}
    \caption{La$^{139}$ NMR linewidth  at low temperatures in La$_{1.88}$Sr$_{0.12}$CuO$_4$. Inset shows the line shape at two different temperatures.  Copyright APS from Ref.~\cite{Yamada2008}. }
    \label{fig:NMR}
\end{figure}

Pairing in a superconductor can also occur through a ``spin-fluctuation" based bosonic mediator as opposed to phonons \cite{Dahm2009}. Detecting anharmonic damping in such bosons requires spin-based experimental probes where the coupling of the probe to the boson occurs through the spin quantum number. Broadly speaking, there are two categories of techniques that can access damping effects in spin based mediators: resonance and magnetization based probes. In resonance based probes -- examples include nuclear magnetic resonance (NMR), nuclear quadrupole resonance (NQR) and  muon spin resonance ($\mu$SR) -- a nucleus or incident muon spin in the sample precesses at its Larmor frequency ($\omega_l$) determined by a combination of an applied and internal magnetic fields. The spin can then decay due to its coupling to the environment, in this case, the fluctuating spins that mediate superconductivity, through hyperfine interactions (for conceptional foundations of the technique, see Ref.~\cite{bloembergen1948relaxation}). Under certain circumstances, either an enhanced decay rate or broadening of the precession linewidth at low temperature can imply a freezing of the spins to due to damping or `glassiness' in the spin fluctuations. This occurs when spin correlation time becomes long enough to be comparable to $\omega_l^{-1}$ and can be indicative of a spin-glass phase with no long-range magnetic order. Existing evidence of glassy spin mediators in the cuprate and iron based superconductors has been established through NMR/NQR~\cite{Curro2013, Hirota2018, Imai2001, Grafe2012, Julien2013, Lin1999, Yamada2008}, $\mu$SR~\cite{Birgeneau1990} and neutron scattering~\cite{Birgeneau1990, Mesot2013}. Fig.~\ref{fig:NMR} shows the broadening of the La$^{139}$ NMR linewidth  in La$_{1.88}$Sr$_{0.12}$CuO$_4$ as seen in Ref.~\cite{Yamada2008}.
Magnetization and magnetic susceptibility are other indicators of freezing of spins and spin-glass physics. Conventional signatures include shift of the ac susceptibility cusp with frequency~\cite{Petrovic2015-Co, Petrovic2015-Ni}, irreversible dc magnetization in the field-cooled and zero field-cooled states~\cite{Aharony1995, Yamada2000,Buechner2012, Petrovic2015-Co, Petrovic2015-Ni, Paulose2010} and a direct measurement of the Edwards-Anderson spin-glass order parameter~\cite{Aharony1995, Yamada2000}. These techniques have been extensively applied in the cuprate~\cite{Aharony1995, Yamada2000} and iron based~\cite{Paulose2010, Buechner2012, Petrovic2015-Co, Petrovic2015-Ni} superconductors where spin fluctuations are thought to play an important role in the pairing mechanism.  A more thorough exposition of the aforementioned topics can be found in Ref.~\cite{Mydosh2015}. 
 \section{Phonon damping from first principles} While the focus of this review is to phenomenologically understand the role of anharmonic damping of the bosonic mediators on superconductivity,  in this section we briefly review existing literature that calculate boson damping/linewidth from first-principles.
 Our objective here is to open up the possibility of integrating bosonic damping into first principles calculations of superconducting properties. Such a scheme could involve incorporating ab-initio data for both the bosonic damping and dispersion relations of real materials into well established routines that evaluate quantities such as Eliashberg functions, coupling constants and critical temperature. 
Most of the focus so far has been on the calculation of anharmonic damping in phonons. 
 The original theories of phonon damping due to anharmonicity~\cite{Fein1962, cowley1965anharmonic, Cowley1965, Cowley1968, Cardona1984} considered anharmonic interactions up to fourth order contributions to the Hamiltonian. Each term in the expansion is associated with harmonic (second-order) and anharmonic (higher order) force constants. The shift of the phonon frequencies and linewidths were evaluated by a diagrammatic perturbation expansion of the self-energy. The broadening of the phonon line was specifically determined using second order perturbation theory of the third order anharmonic term in the expansion of the total energy. This approach was applied to Raman linewidths of Si, Ge and $\alpha$-Sn~\cite{Cardona1984} and it was argued that a combination of optical and acoustical phonons were the key decay channels that contributed to the linewidths. More recently, Green's function based methods have been advanced to study the role of anharmonic damping and other quantum effects on the phonon density of states~\cite{Singh2023} and electron-phonon couplings~\cite{Errea2023}.

 The simplest incorporation of theoretically determined linewidths into first principles is to evaluate the harmonic and anharmonic force constants \textit{ab initio}.  For example, the matrix elements of the anharmonic tensor that contribute to the linewidths are third-order differentials of the total free energy for a single unit cell with respect to the phonon displacement amplitudes. These can then be obtained via density functional perturbation theory~\cite{Vigneron1989, Debernardi1994, Debernardi1995, Debernardi1998, Debernardi1999}. This approach was applied to optical phonons at the zone center in Ge, Si, and C where the temperature and pressure dependencies of the linewidth were calculated~\cite{Debernardi1995,Debernardi1994} and shown to be in good agreement with experiments. Similarly, longitudinal and transverse linewidths of zinc-blende semiconductors such as AlAs, GaAs, InP and GaP were determined~\cite{Debernardi1998} and the temperature dependence of the damping was shown to be consistent with Raman data. More recently, density functional second order perturbation theory was applied to noble metals~\cite{Fultz2011} Cu, Ag and Au, and has been used to understand thermal conductivity and phonon linewidths  in the dichalcogenide MoS$_2$~\cite{Mingo2013}. A similar approach was used to show that dynamical phonon anomalies (beyond the Born-Oppenheimer approximation) can considerably modify the electron-phonon coupling strength $\lambda$ and transition temperature T$_c$ in conventional superconductors~\cite{Dovko2023}. First principles calculations to evaluate accurate interatomic forces was also applied to study the thermodynamics of crystals at finite temperature taking into account anharmonic effects. Termed as ``self-consistent ab initio lattice dynamics (SCAILD)"~\cite{Rudin2008}, the method was employed to understand the stability of several body-centered cubic metals whose lattice structure was unstable at low temperature but stabilized at higher temperatures. 

An alternate approach toward first principle computation of linewidths is a simplified version of the Car-Parinello scheme~\cite{CP1985, CP1988}. The Car-Parinello method  unifies density function theory and molecular dynamics simulations to provide an accurate description of the inter-atomic forces, ground state and finite temperature properties (like energy shifts and linewidths) of quantum mechanical systems. However, the method is computationally intensive when applied to real materials. Refs~\cite{Ho1989-1, Ho1989-2, Ho1990} simplified the scheme by calculating free energies and inter-atomic forces by combining molecular dynamics with an empirical tight-binding method rather than density functional theory. This enabled an approximate but efficient way of analysing lattice and electronic properties. The method was used to study temperature dependence of anharmonic frequency shifts and linewidths in Si and diamond~\cite{Ho1989-1, Ho1989-2, Ho1990}, and was shown to be in good agreement with data. A molecular dynamics based approach was also implemented to understand the stability of body-centered cubic lattice phases of Li and Zr at high temperatures~\cite{Simak2011}.   \par 

  \begin{figure}
    \centering
    \includegraphics[width=1\linewidth]{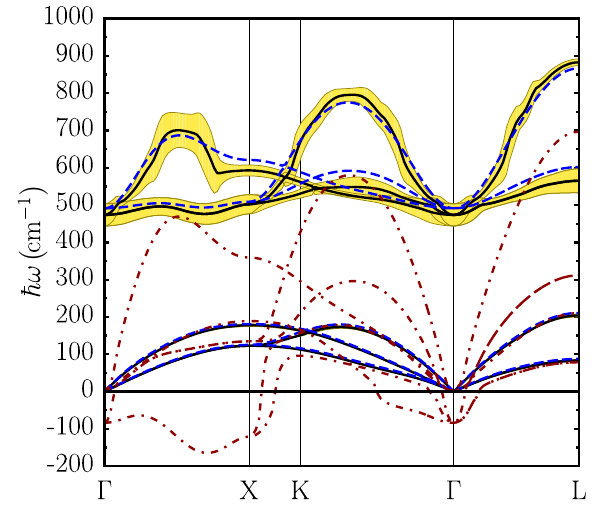}
    \caption{Phononic dispersions and linewidths for palladium hydride at $295$ K. Red (blue) dashed line is the spectrum from harmonic approximation (SSCHA). The shaded yellow region is the calculated linewidth.  Copyright APS from Ref.~\cite{Mauri2015-2}. }
    \label{fig:SSCHA}
\end{figure}

 First-principles evaluation of anharmonic phonon properties has made rapid progress in recent years since the discovery of hydride superconductors. In the hydrides, anharmonicity is known to be substantial and standard perturbative approaches fail \cite{Mauri2015}. Typically, variational approaches such as the self-consistent harmonic approximation (SCHA)~\cite{Born, Hooton1955, Koehler1966} are employed in non-perturbative settings. In the last couple of years, the method has also been generalized to include time dependent effects to simulate nuclear dynamics~\cite{Mauri2023,Mauri2021}. However, this approach is computationally intensive  and a stochastic version of the SCHA (called the SSCHA) has been explored~\cite{Mauri2013, Mauri2014, Mauri2015, Mauri2015-2, Mauri2016} to determine anharmonic free energy, thermal transport and superconducting properties. In the case of palladium and platinum hydrides, it was shown that phononic spectra are strongly renormalized by anharmonicity and harmonic approximations overestimate the superconducting transition temperature~\cite{Mauri2013, Mauri2014}. The SSCHA allows computation of anharmonic phonon linewidths arising from phonon-phonon interactions. Fig.~\ref{fig:SSCHA} shows the full width half maximum of palladium hydride phonon spectrum obtained from SSCHA~\cite{Mauri2015-2}.  Alternatively, one can adopt the self-consistent phonon theory (SCP) with anharmonic force constants, see \textit{e.g.} \cite{Tadano,Tadano1,Tadano2} for recent developments and the current state of the art. These techniques directly stem from the original work of Born and Hooton \cite{Born,Klein1972}. A brief review of various first principle methods for treating anharmonicity and phonon lifetimes can be found in Ref.~\cite{Tadano}. Despite these attempts, currently there are no systematic studies that take into account anharmonic damping effects to examine superconducting properties from first principles, and efforts in this direction are much needed.

 \section{Damped bosons: Minimal Theory of Superconductivity}
 In this Section, we would like to understand how the superconducting properties, and specially the critical temperature $T_c$, are affected by the low-energy parameters appearing in the dispersion relations Eqs.\eqref{dispacu}-\eqref{opt}. In order to make the analysis more concrete, in the following we will explicitly identify the bosonic mediator with acoustic and optical phononic modes.
 \subsection{BCS Theory: Acoustic phonons}\label{BCSsection}
The discussion in the next two subsections follows Ref.~\cite{Setty2020}.  We first consider the situation in which the bosonic mediators are acoustic phonons that obey the simple dispersion relation in Eq.\eqref{dispacu}. Moreover, in this Section, we will limit ourselves to standard BCS superconductors described BCS theory (see \cite{Carbotte2008} for a comprehensive review). SWe start our discussion from the Green's function expressed in Eq.\eqref{green}, which allows us to re-write the phonon propagator as:
\beq 
\Pi(\Omega_n,k)= \mathcal{G}(i \Omega_n,k) = \frac{1}{v^{2}k^2 +\Omega_n^2+D k^2|\,\Omega_n|}.
\eeq
Here, $\Omega_n= 2 \pi n T$ correspond to the bosonic Matsubara frequencies where $T$ is the temperature of the system and $n$ an integer index which serves as label. 
The phonon damping, or linewidth, which appears in the last term in the denominator of the above expression must be positive, due to the analytical properties of the response functions involved in the quantum theory of dissipative systems, \textit{cfr.}, Refs.\cite{Leggett1983,Weiss}.

\begin{figure}[h!]
\includegraphics[width= 0.8\linewidth]{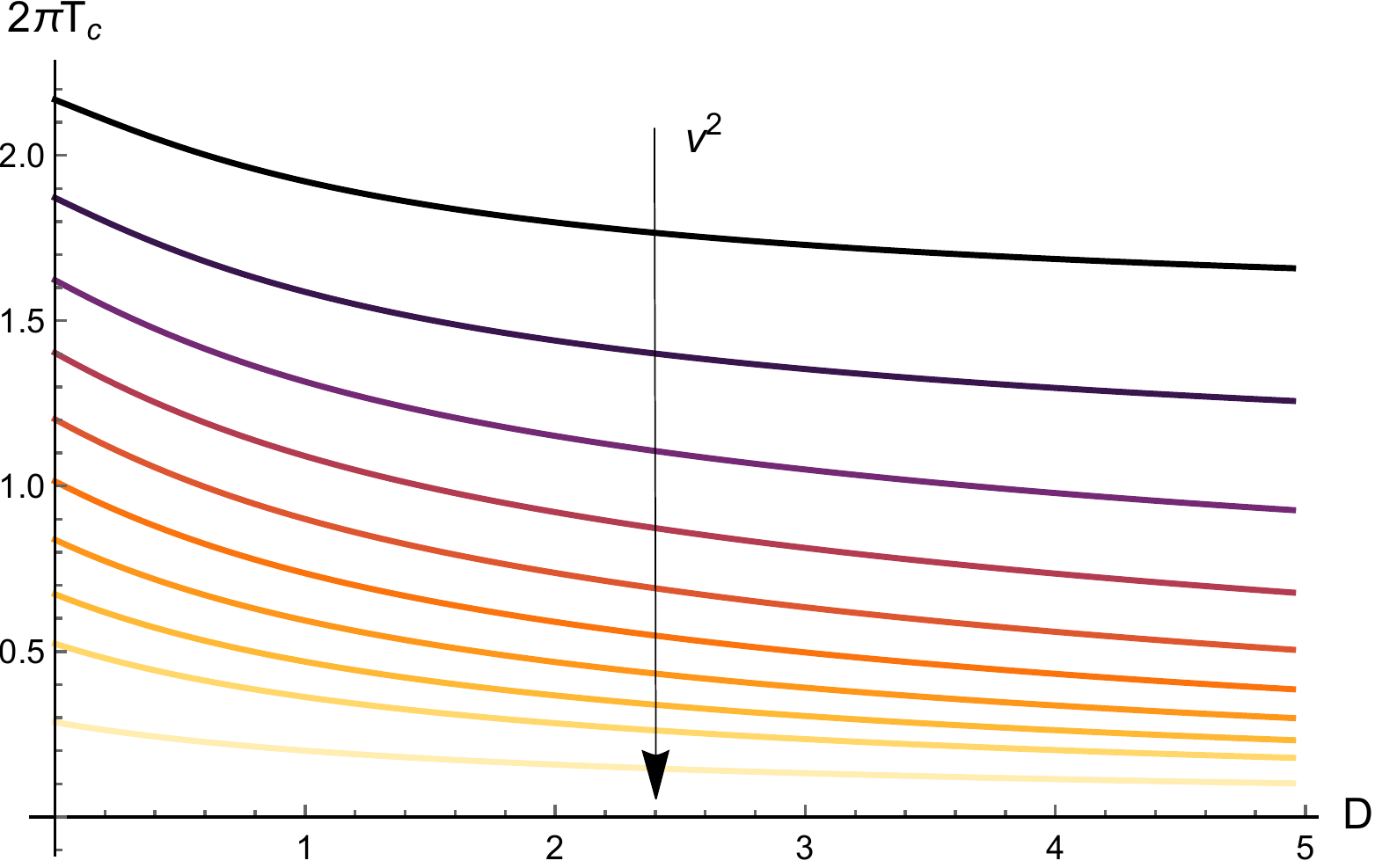}
\caption{Superconductivity mediated by acoustic phonons. The critical temperature $T_c$ as a function of the damping constant $D$ for  several values of the velocity of acoustic phonons
 $v \in [0.8, 1.8]$ (from black to yellow). Copyright APS from Ref.~\cite{PhysRevB.106.139901}.   
} \label{TcVsD} 
\end{figure}

Common algebraic manipulations \cite{Carbotte2008,Kleinert} yield to the superconducting gap equation,
\beq \nonumber
\Delta(i\omega_n, k) &=& \frac{g^2}{\beta ~V} \sum_{q, \omega_m} \frac{\Delta(i\omega_m, k+q) \,\Pi(k, i\omega_n - i \omega_m)}{\omega_m^2 + \xi_{k+q}^2 + \Delta(i\omega_m, k+q)^2}\,,\\
\label{Sum-GapEqn}
\eeq
where $g$ is the coupling that quantifies the attractive pairing interaction, $V$ the volume and $\beta \equiv 1/T$. Moreover, $\xi_k\equiv k^2-\mu$ is the free electron dispersion in presence of a chemical potential $\mu$. For simplicity, we set the electron mass to $2 m_e=1$ and work in reduced units. Finally, we assume the gap  $\Delta(i\omega_n, k)=\Delta$ to be independent of the frequency and the wave-vector. Then, we replace the sum over the wave-vector $k$ with a continuous integration over the energy $\xi$,
\begin{equation}
    \frac{1}{V}\sum_{\textbf{q}} \rightarrow \frac{1}{(2 \pi)^d}\int d^d\textbf{q} \rightarrow \int N(\xi)d\xi\,,
\end{equation}
where $N(\xi)$ is the density of states at energy $\xi$. We then assume a constant density of state $N(\xi)\approx N(0)$. All in all, the superconducting critical temperature $T_c$ can be readily obtained by imposing that the superconducting gap vanishes, $\Delta =0$. The behavior of the critical temperature $T_c$ is plotted in Fig.~\ref{TcVsD} as a function of the anharmonic damping parameter $D$.  The figure illustrates that $T_c$ always decreases monotonically. Physically, this implies that anharmonicity, $\propto D$, is always detrimental for the onset of superconductivity under these assumptions (see \cite{PhysRevB.106.139901} for more details).\\

 \subsection{BCS Theory: Optical phonons}
 Until now, we have mainly considered the case in which the electronic pairing is mediated by the interaction with acoustic phonons. In this case, because of their acoustic nature the damping is a quadratic function of the wave-vector $k$,\textit{i.e.}, $\Gamma \sim k^{2}$ (Akhiezer mechanism). Here, we consider the alternative scenario in which the ``glue'' is provided by optical phonons. In this case, the damping coefficient, or phonon lifetime, is independent of the wave vector, as derived using perturbation theory by Klemens~\cite{Klemens}. Moreover, the lifetime of the optical phonons is controlled by their decay into acoustic phonons and  ultimately by the Gr{\"u}neisen constant squared.

\begin{figure}[hbt]
\includegraphics[width= 0.6\linewidth]{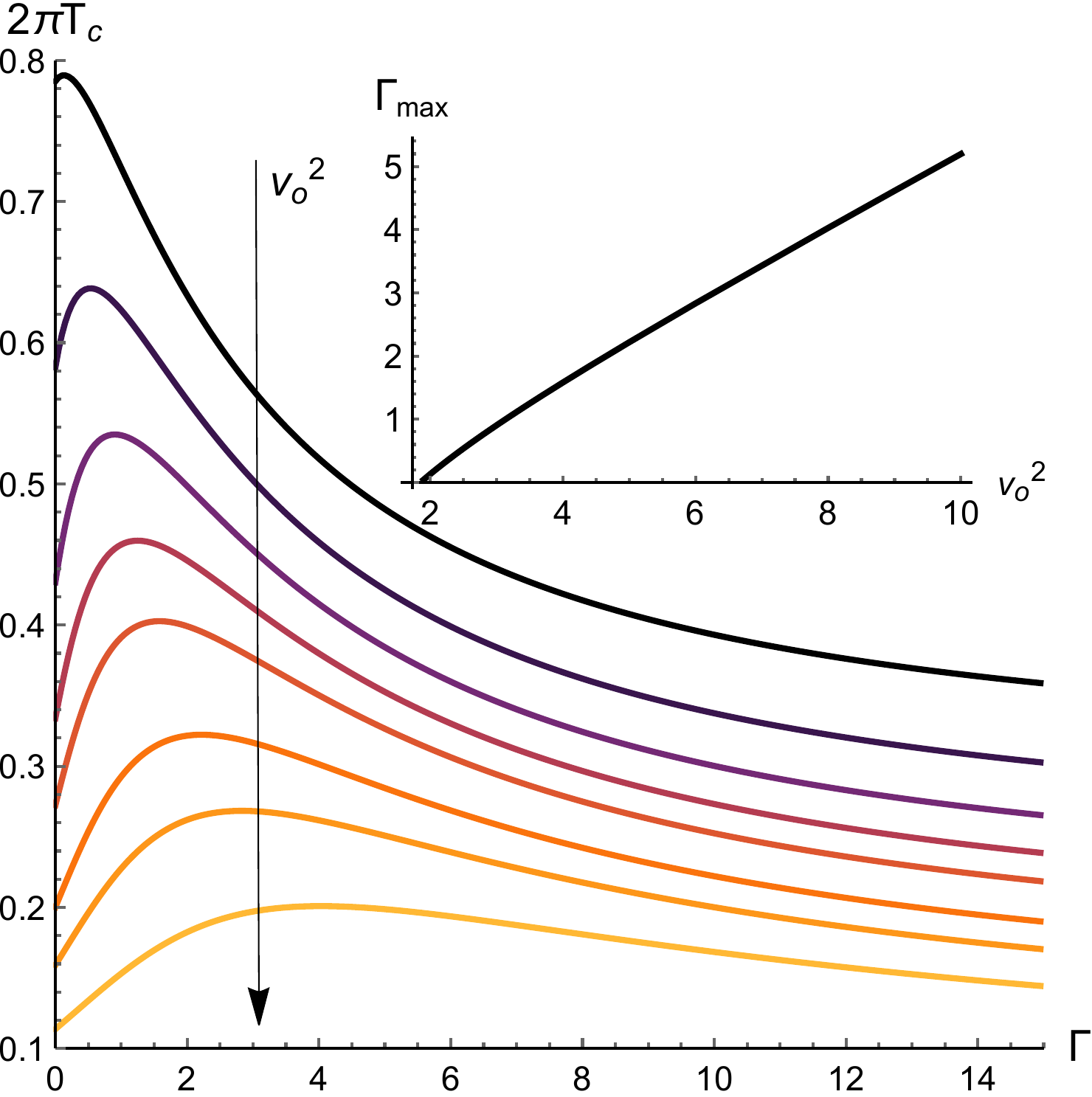}

\vspace{0.4cm}

\includegraphics[width= 0.6\linewidth]{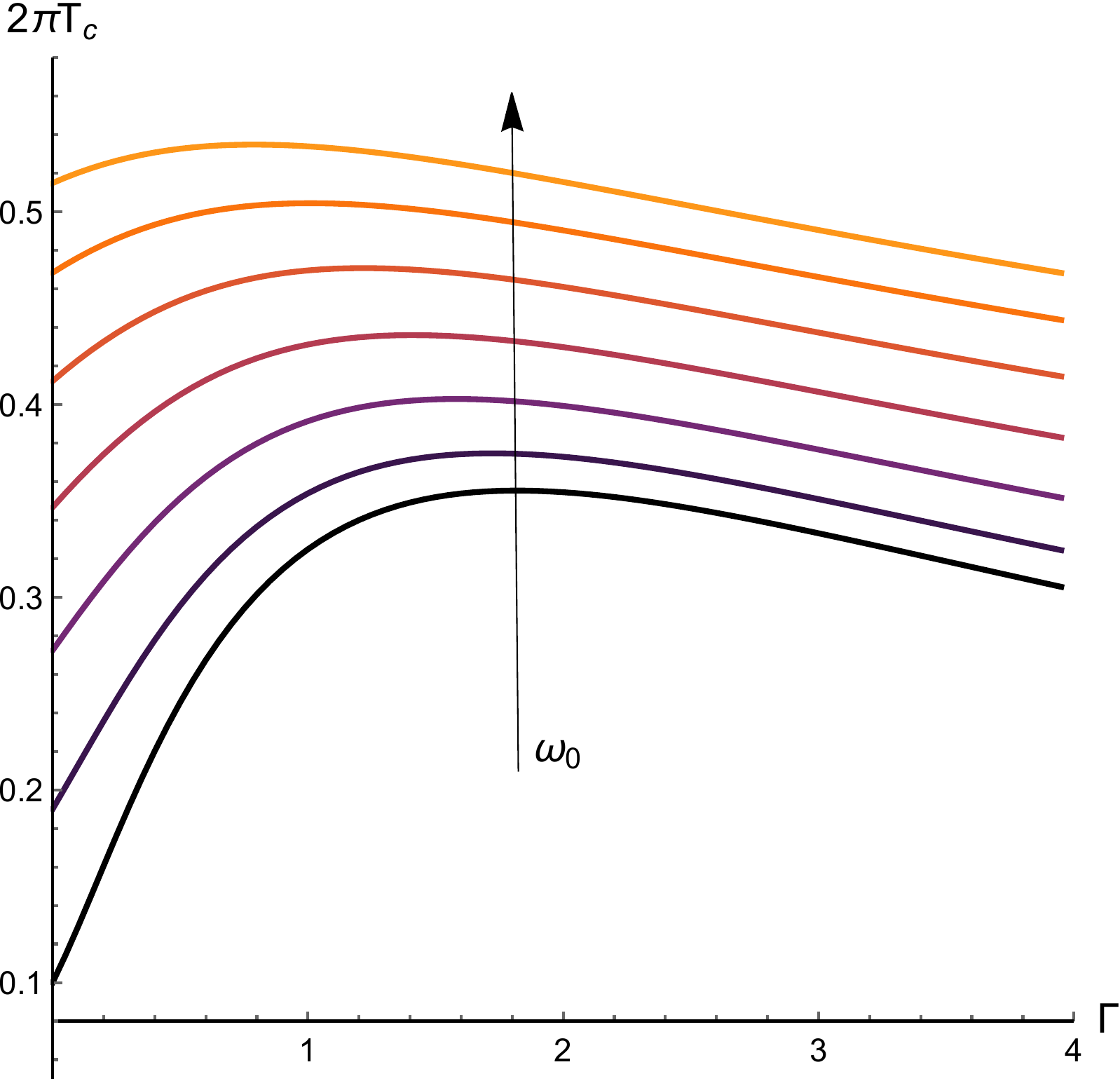}
\caption{Optical phonon mediated superconductivity. Plots of critical temperature as a function of damping for various $v_o$ (top) and optical gap/mass (bottom). \textbf{(Top)} $v_o^2 \in \left[2,8\right]$ (from black to yellow), where $v_o^{2}=2 \omega_0 \alpha$, and the optical mass is fixed to $\omega_0=0.3$. \textbf{(Bottom)} $\omega_0 \in \left[0.1,0.7\right]$ (from black to yellow) with a fixed value $v_o^2=4$. Copyright APS from Ref.~\cite{PhysRevB.106.139901}.
}\label{CPEffect}
\end{figure}

For the optical phonon modes, we assume the following dispersion relation
\begin{equation}
\Omega_{\text{opt}}(k)=\omega_0 +\,\alpha\, k^{2}.
\end{equation}
where $\omega_0$ is the optical phonon mass and $\alpha$ is curvature of the dispersion. We further choose a damping independent of wave vector,  which we denote by $\Gamma$. With these assumptions, the  bosonic propagator takes the following form
\begin{equation}
\Pi(i \Omega_n,k) = \frac{1}{\left[\omega_{0}^2 +2 \,\omega_0\,\alpha \,k^{2}+\mathcal{O}(k^4)\,\right] +\Omega_n^2+\Gamma\,|\Omega_n|}.
\end{equation}
We now implement the propagator $\Pi(i \Omega_n,k)$ above into the gap equation. The theoretical predictions for $T_c$ as a function of anharmonic damping constant $\Gamma$ that we obtain are shown in 
in Fig.\ref{CPEffect}.

Our predictions demonstrate that low and moderate anharmonic phonon damping can lead to an 
enhancement of the critical temperature. 
The rise of $T_c$ is followed by a peak for optimal damping and then subsequently a decrease for very large values of anharmonic damping. Therefore, it is evident that  anharmonic damping can lead to a substantial increase in $T_c$ for a range of mass values. This behavior must be contrasted with the case of acoustic phonons where $T_c$ is monotonically suppressed. 
Furthermore, as shown in the lower panel of Fig.\ref{CPEffect}, our model predicts that the damping-induced enhancement, and the peak, become larger upon decreasing the optical phonon energy gap $\omega_0$. 
Finally, one can also examine the role of the curvature coefficient, $\alpha$, in the optical dispersion relation on the transition temperature. It affects the percent of enhancement as well as the peak peak value -- both become larger as the coefficient $\alpha$ becomes smaller.  Hence approaching a flat optical dispersion, typically seen in DFT computations of optical phonons in hydride materials~\cite{Pickard2015}, is favorable to the absolute value of $T_c$; however, the enhancement effect is reduced in the process.
To understand the increase in $T_c$, we observe that phonon dispersion and anharmonic damping behave in such a way as to superpose bosons at different energy scales. The superposition acts to combine various phonons with different energies coherently to increase the superconducting transition temperature. This occurs because, mathematically, the $k$-dependence in the propagator is integrated out in the gap equation. Such an integration combines low and high energy phonons coherently and, thereby, increasing the electron-phonon coupling effectively.
In the opposite limit when the linewidth dominates the dispersion spectrum, the phonons are no longer able to provide a sufficiently strong pairing for the electrons. Thus there is eventually a suppression of $T_c$ for sufficiently large $\Gamma$ as demonstrated in Fig.\ref{CPEffect}.

  \subsection{Eliashberg Theory for damped phonons}
  \label{thel}
  In this Section, we consider a different theoretical approach in which the superconducting transition is treated by means of Eliashberg theory \cite{BSZ2020,marsiglio2008electron}. The discussion in this subsection closely follows Ref.~\cite{BSZ2020}.  More concretely, as mediators, we consider acoustic phonons with quadratic attenuation constant whose Green's function is parameterized as usual:
  \begin{equation}
    G(\omega,k)\,=\,\frac{1}{\omega^2\,-\,\Omega^{2}(k)\,+\,i\,\omega\,\Gamma(k)}\label{e1},
\end{equation}
with propagating term given by $\Omega^{2}= v^2\,k^2\,$ and the attenuation constant by $\Gamma=D\,k^2$. From Eq.\eqref{e1}, we can derive the corresponding spectral function which is given by \cite{BSZ2020}:
\begin{equation}
    \mathcal{B}(\omega,k)\,=\,\frac{\omega\,\Gamma(k)}{\pi[\left(\omega^2\,-\,\Omega^2(k)\right)^2\,+\,\omega^2\,\Gamma^2(k)]}\,.
\end{equation}
We can then express Eliashberg spectral function in the following form~\cite{marsiglio2008electron}:
\begin{equation}
    \alpha^2\,F(\Vec{k},\Vec{k}',\omega)\,\equiv\,\mathcal{N}(\mu)\,|g_{\Vec{k},\Vec{k}'}|^2\,\mathcal{B}(\Vec{k}-\Vec{k}',\omega).
\end{equation}
In the formula above, $\mathcal{N}(\mu)$ represents the electronic density of states computed at $\mu$ (chemical potential). Additionally, $g_{\Vec{k},\Vec{k}'}$ is the electron-phonon matrix element. Following the steps in \cite{marsiglio2008electron}, spectral function that is averaged over the Fermi surface reads:
\begin{equation}
    \alpha^2\,F(\omega)=\frac{1}{\mathcal{N}(\mu)^2}\,\sum_{\Vec{k},\Vec{k}'}\,\alpha^2\,F(\Vec{k},\Vec{k}',\omega)\,\delta(\epsilon_{\Vec{k}}\,-\,\mu)\,\delta(\epsilon_{\Vec{k}'}\,-\,\mu)\, .\label{ee}
\end{equation}
For simplicity, we take the matrix elements to be constant in wave-vector, \textit{i.e.}, $g_{\Vec{k},\Vec{k}'}\equiv \mathrm{g}$. In this way, the previous equation takes the simplified form:
\begin{equation}
   \alpha^2\,F(\omega)=\frac{\mathrm{g}^2}{\mathcal{N}(\mu)}\,\sum_{\Vec{k},{\Vec{k}}'}\,\mathcal{B}(\Vec{k}-{\Vec{k}}',\omega)\,\delta(\Vec{k}^2\,-\,\mu)\,\delta({{\Vec{k'}}}^2\,-\,\mu)\,.
\end{equation}
In the expression above, we have assumed the typical electronic band of the form $\epsilon_{\Vec{k}}=\Vec{k}^2$. To make further progress, we convert the previous sum into a 2D integral using the relation $
    \sum_{\Vec{k}}=\,\frac{V_2}{(2\pi)^2}\,\int \,k\, dk\, d\phi_k $
with the wave-vector amplitude $k \in [0,\infty]$ and $\phi_k\in [0,2\pi]$. Spatial isotropy dictates that $\mathcal{B}(\Vec{k}-\Vec{k}',\omega)$ depends only on the difference $( \Vec{k}\,-\Vec{k}')^2$, which can be expressed in polar coordinates as 
\begin{equation}
    ( \Vec{k}\,-\Vec{k}')^2\,=\,k^2\,+\,{k'}^2\,-\,2\,k\,k'\,\cos(\phi_k\,-\,\phi_{k'})\,. 
\end{equation}
All in all, we can perform the above integral and obtain the final result \cite{BSZ2020}:
\begin{align}
    &\alpha^2\,F(\omega)\,=\,\frac{\mathrm{g}^2}{4\,(2\pi)^4\,\mathrm{N}}\,\int  \mathcal{B}(X^2,\omega)\,d\phi_k\,d\phi_{k'} \label{aa}\\
    &\mathcal{B}(X^2,\omega)\,=\, \frac{\omega\,D\,X^2}{\pi\left(\omega^2\,-\,v^2\,X^2\right)^2\,+\,\omega^2\,D^2\,X^4}\label{eq9}\\
    &  X^2\,\equiv\,2\,\mu\,\left(1\,-\,\cos(\phi_k\,-\,\phi_{k'})\right) \label{eq10}
\end{align}
where the electronic density of states is assumed to be constant, $\mathcal{N}(\mu)=\mathrm{N}$. This final integral in Eq.\eqref{aa} can be performed numerically.\\

At this point, we can use the standard definition for the electron-phonon mass enhancement parameter,
\begin{equation}
    \lambda(v,D)\,=\,2\,\int_0^\infty\, \frac{\alpha^2\,F(\omega)}{\omega}\,d\omega\,,\label{eq11}
\end{equation}
determining the effective (dimensionless) strength of the electron-phonon interactions.
In order to estimate the critical temperature $T_{c}$, we use the Allen-Dynes formula ~\cite{PhysRevB.12.905} given by:
\begin{equation}
      T_c\,=\,\frac{f_1\,f_2\,\omega_{log}}{1.2}\,\exp\left(-\frac{1.04\,(1+\lambda)}{\lambda-u^\star\,-\,0.62\,\lambda\,u^\star}\right)\label{allenformula}
\end{equation}
where 
\begin{equation}
\omega_{log}=\exp \left(\frac{2}{\lambda}\int_{0}^{\infty} d\omega \frac{\alpha^{2}F(\omega)}{\omega} \ln \omega \right) \label{eq13}
\end{equation}
represents the characteristic energy scale of phonons for pairing in the strong-coupling limit, while $f_1$, $f_2$ are semi-empirical correction factors, as defined in ~\cite{PhysRevB.12.905}.
The parameter $u^\star$ encodes the strength of the Coulomb interactions and it  is determined experimentally and tabulated in the literature for various materials; we will take it as an external input from tabulated literature data. That said, all the SC properties are determined by the shape of the spectral function $\alpha^2 F(\omega)$.

\begin{figure}
    \centering
    \includegraphics[width=0.85\linewidth]{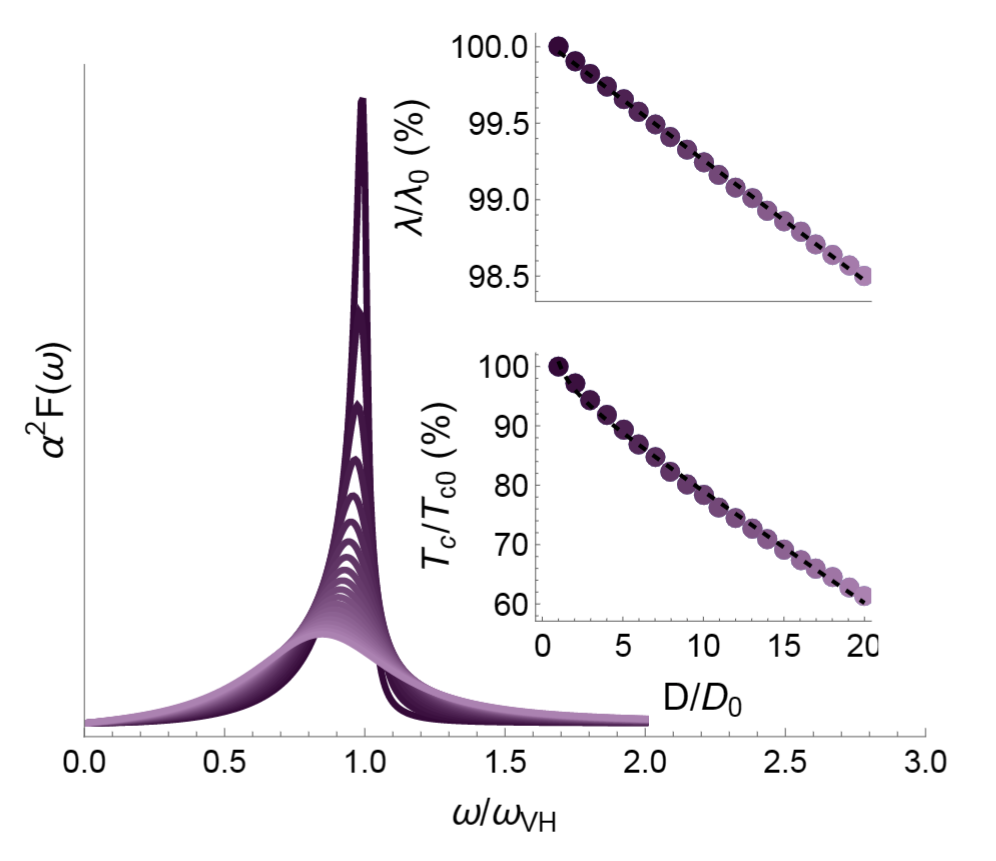}
    \caption{The Eliashberg function $\alpha^2 F(\omega)$ corresponding to the mediator dispersion relation in Eq.\eqref{ioio}. The insets show the coupling constant $\lambda$ and $T_c$ as a function of the damping parameter $D$. $\lambda_0$ and $T_{c,0}$ are the values at the smallest damping $D=D_0=100$. In this figure, the other parameters are fixed to $v=5000$, $k_{\mathrm{VH}}=1$, $k_F=1/2$, $g_k^2=C (vk)^2$, $C=0.03$ and $\mu^*=0.1$. Copyright IOP from Ref.~\cite{jiang2022sharp}.}
    \label{fig:wq}
\end{figure}

As a concrete application of this framework, let us consider as a bosonic mediator an acoustic phonon described by the following choice,
\begin{equation}\label{ioio}
    \Omega(k)=v k-\frac{v}{2 k_{\text{VH}}}k^2\,,\quad \Gamma(k)=D k^2\,,
\end{equation}
where $v$ is the speed of sound, $D$ the attenuation constant and $k_{\text{VH}}$ the location of the Van-Hove singularity. In Fig.\ref{fig:wq}, we show the results for this choice of mediators. The Eliashberg function $\alpha^2 F(\omega)$ shows a clear peak which broaden upon increasing the attenuation constant $D$. Both the effective electron-phonon coupling $\lambda$ and the critical temperature $T_c$ decrease monotonically upon increasing the attenuation constant of the acoustic phonon which mediates the pairing. In other words, one finds that, in the case of acoustic phonons, the anharmonic damping is detrimental to superconductivity.

  \subsection{BCS Theory: Glassy spins}
  Randomness, when taken into account collectively through electron correlation effects yields interesting phases of matter~\cite{Ramakrishnan1985}. The spin glass (SG) phase forms one such example which has been thoroughly studied as well as observed in the phase diagrams of correlated electronic systems~\cite{Aharony1995, Yamada2000, Ferretti2001, Imai2001, Julien2003, Birgeneau1990, Yamada2008, Hirota2018, Lin1999, Grafe2012, Julien2013, Curro2013, Mesot2013, Petrovic2015-Co, Petrovic2015-Ni, Buechner2012, Paulose2010}. The phenomenology of the SG is remarkable~\cite{Mydosh2015}. Typically, the SG phase is characterized by aging, linear in temperature of the AC susceptibility peak, hysteretic effects in the DC magnetization and a cusp in the thermodynamic specific heat, to name a few. 
  On the theoretical side, SG order phase occurs when the spin average on each lattice site is non-zero, but vanishes when spatially averaged over the entire lattice~\cite{Ye1995}.  For the purposes of this discussion, we note a key property of the temporal dependence $\tau$ of spin correlation function at the SG critical point: it follows a power law of the form~\cite{Moore1985, Huse1993, Ye1995, Phillips1999}
\beq
\Pi(\tau) \equiv \bigg[\langle S_{i\mu}(\tau)S_{i\mu}(0)\rangle \bigg] \sim \frac{1}{\tau^2}.
\label{SGCorrelator}
\eeq 
 We define $S_{i\mu}$ as the spin at site $i$ with its $\mu-$th component, and the square and angular brackets denote the site and thermal averages respectively. In frequency space, the the correlation function behaves as $\Pi(\omega)\sim |\omega|$, \textit{i.e.}, it is linear in frequency indicating that dissipation is a necessary (but not sufficient) condition for a SG. \par
 Spin fluctuations have been studied extensively as potential mediators of  superconductivity~\cite{Ketterson2008, hirschfeld2011gap}. In the proximity of a SG phase, correlators of the form appearing in Eq.~\ref{SGCorrelator} modify the spin fluctuation propagator~\cite{Setty2019}. Since the dissipative component arises from randomness that is exclusively a property of the spin sector, it constitutes an anharmonicity of the spin mediator. To see how this occurs, we write the model for the total bosonic propagator by additing the dissipative (anharmonic) contribution from Eq.~\ref{SGCorrelator}. As a result, the total action consists of a free term $S_0[\Psi, \Psi^*] $ and a dissipative term $ S_{\text{diss}}[\Psi, \Psi^*]$. These are given by
\beq \nonumber
S[\Psi, \Psi^*] &=& S_0[\Psi, \Psi^*] + S_{dis}[\Psi, \Psi^*] \\ \nonumber
S_0[\Psi, \Psi^*] &=& \int d^d \bs r d\tau \bigg[ \kappa |\nabla \Psi(\bs r, \tau)|^2 + |\partial_{\tau} \Psi(\bs r, \tau)|^2 \\\nonumber
&& + M^2 |\Psi(\bs r, \tau)|^2  \bigg],\\ \nonumber
S_{\text{diss}}[\Psi, \Psi^*] &=& \sum_{\bs k, \omega_n} (2\eta~|\omega_n|)|\Psi(\bs k, \omega_n)|^2. 
\eeq
Here $\Psi$ is the bosonic field, $\omega_n$ is the Matsubara frequency, $\eta$ is the dissipation (anharmonicity) parameter, $\kappa$ is the  energy scale of the bosonic velocity (or spatial stiffness), and the parameter $M^2$ is the square mass that is proportional to the inverse correlation length. 
The bosonic propagator $\Pi(\bs k, i \omega_{n} - i \omega_m)$ for the action $S[\Psi, \Psi^*]$ takes the form 
    \beq \nonumber
    \Pi(\bs k, i \omega_{n}) = \frac{\alpha}{\kappa k^2 + \omega_n^2 + 2 \eta |\omega_n| + M^2 }.
       \eeq
Here $k = |\bs k|$ and  $\alpha$ is a constant with dimensions of energy (see for example~\cite{Ketterson2008}). \par

We can now substitute $\Pi(\bs k, i \omega_{n} - i \omega_m)$ into Eq.~\ref{Sum-GapEqn}. We assume a frequency independent and isotropic $s$-wave gap (henceforth denoted by $\Delta$) and perform the momentum and Matsubara summations. The equation determining $T_c$ (setting $\Delta =0$) then becomes~\cite{Setty2019}
\beq \nonumber
 1 &=& - \lambda \Bigg[ \frac{\psi\left(\frac{1}{2} + \frac{\eta' -  i \kappa}{2 \pi T_c}\right)}{2\left(\eta' - i \kappa \right)^2} +  \frac{\psi\left(\frac{1}{2} + \frac{\eta' + i \kappa}{2 \pi T_c}\right)}{2\left(\eta' + i \kappa \right)^2} \\
 && + \frac{\kappa^2 - \eta'^2}{(\kappa^2 + \eta'^2)^2} \psi\left( \frac{1}{2} \right) - \frac{\pi^2 \eta'}{4 \pi T_c (\eta'^2 + \kappa^2)}\Bigg]\,,
 \label{Non-Local-GapEqn}
 \eeq
where $\eta' \equiv 2 \eta$ and $\psi(x)$ is the digamma function. The Eq.~\ref{Non-Local-GapEqn} can be solved numerically to study the effect of $\eta$ on $T_c$. As will be further elucidated below, $T_c$ follows a non-monotonic behavior with $\eta$ where the optimal value is set by the stiffness $\kappa$. A potentially interesting limiting case is that of $\kappa \rightarrow 0$, in which case the gap equation reduces to 
 \beq \nonumber
 1 &=& \frac{\lambda  \left( \eta - i \bar{M}\right)^{-1}}{2 i \bar{M}} \bigg[ \psi\left(\frac{1}{2} + \frac{\eta}{2 \pi T_c} - i \frac{\bar M}{ 2\pi T_c} \right) - \psi\left(\frac{1}{2}\right) \bigg] \\
 && +~~~~ c.c ,
  \label{Local-GapEqn}
 \eeq
 where $\bar{M} \equiv \sqrt{M^2 -\eta^2}$. 
As we will see below, in this limit, $T_c$ monotonically decreases with $\eta$. 
 \subsection{Optical soft mode instabilities and structural transitions}
 \label{secsoft}
 Soft phonon modes appear near structural transitions in which a higher-symmetry crystal structure transforms into a lower-symmetry one \cite{venkataraman1979soft}. Typical examples of this sort are ferroelectric and ferroelastic transitions \cite{xu2013ferroelectric,gonzalo2006effective}.
Here, we address the question of how the appearance of soft mode instabilities in a metallic state might affect the critical temperature of a near superconducting transition. For simplicity, we will focus on the usage of BCS theory and on the situation in which the soft mode is a optical excitation whose dynamics is described by Eq.\eqref{optdisp}. Solving Eq.\eqref{optdisp}, we obtain the simple dispersion relation:
 \begin{equation}
     \omega=\pm \sqrt{\omega_0^2-\frac{\Gamma}{4}}-\frac{i}{2}\Gamma
 \end{equation}
 from which we can identify the real part as the renormalized energy and the imaginary part as the inverse lifetime $\tau^{-1}$. In the limit of $\Gamma \rightarrow 0$, the relation above coincides with the Einstein approximation $\omega=\omega_0$. In order to continue, and as we will see later to make contact with realistic materials, we assume a soft mode instability in which the energy of the soft mode is well described (at least close enough to the critical point) by the mean-field Curie-Weiss law:
 \begin{equation}\label{real1}
 \omega_0^2\sim |n-n_c|,
 \end{equation}
 with $n$ an external parameter driving the instability \cite{cowley2012soft}. By doing so, the dynamics of the low-energy critical modes is defined by the following dispersion relations:
 \begin{equation}\label{mm}
     \omega=\frac{1}{2} \left(\pm\sqrt{4 | n-n_c| -\Gamma ^2}-i \Gamma \right)\,.
 \end{equation}
 \begin{figure}
     \centering
     \includegraphics[width=0.8\linewidth]{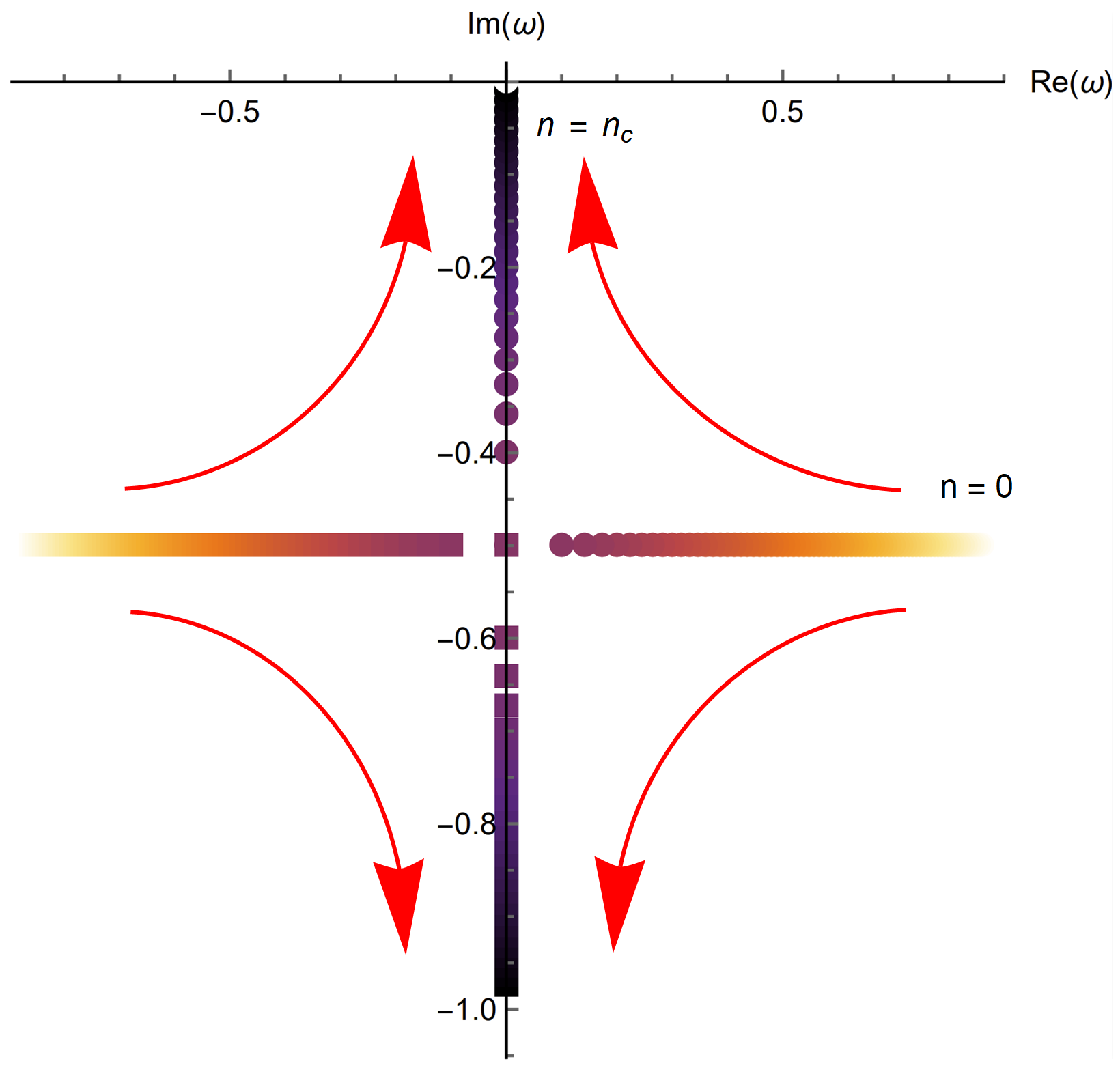}
     \caption{The dynamics of the critical modes from Eq.\eqref{mm} as a function of $n-n_c$. Here, $\Gamma=1$. From white color to black color we change the control parameter from $n=0$ to $n=n_c$ as the red arrows indicate.}
     \label{fig:modes}
 \end{figure}
 
 As evident from Eq.\eqref{mm}, and as shown in Fig.\ref{fig:modes}, at the critical point one of the two modes becomes strongly overdamped, $\omega=-i \Gamma$, while the other approaches the origin of the complex plane, $\omega=0$, moving along the imaginary axes. This second mode is the mode responsible for the instability at $n=n_c$.\\

\begin{figure}[h]
     \centering
     \includegraphics[width=0.8\linewidth]{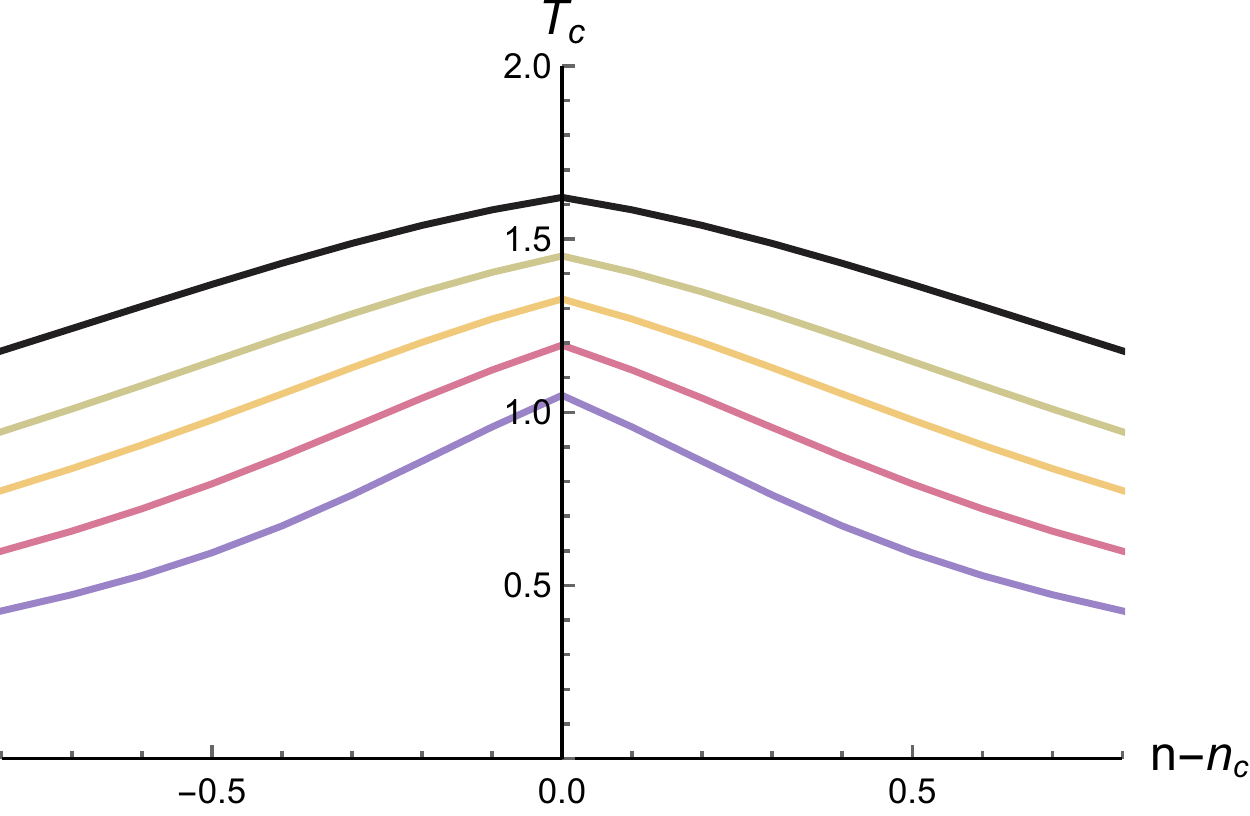}
     \caption{The critical temperature as a function of $n-n_c$ for $v^2=\mu=1$ and increasing the linewidth $\Gamma$ from black to light blue. Copyright APS from Ref.~\cite{PhysRevB.105.L020506}.}
     \label{fig:boh}
 \end{figure}

Using the framework described in the previous section, we can then compute the critical temperature $T_c$ as a function of the distance from the critical point, $n-n_c$ (see \cite{Setty2022,PhysRevB.105.L020506} for details). The critical temperature displays a dome shaped behavior centered at the critical point which is shown in Fig.\ref{fig:boh} for different values of the damping parameter $\Gamma$. A larger value of $\Gamma$ corresponds to a more pronounced maximum at $n=n_c$. In summary, this simple model predicts the appearance of a dome shaped critical temperature $T_c$ which is maximized at the location of the critical point \cite{Setty2022}. Such a result provides a viable explanation (see \cite{PhysRevLett.115.247002} for a different explanation based on the concept of quantum criticality) for the superconducting dome experimentally observed in various ferroelectric materials such as SrTiO$_3$ \cite{PhysRev.163.380} (see Section \ref{pala}).

\subsection{Kohn-like soft phonon instabilities}
Phonon softening is a general phenomenon in condensed matter physics which is not restricted to structural phase transitions. A particularly interesting scenario is that associated with the formation of charge order in metallic states, in which softening emerges in the form of  ``Kohn anomaly'' \cite{PhysRevLett.2.393}. Charge correlations soften the energy of the acoustic phonons whose dispersion presents a pronounced dip at a finite value of the wave-vector. The frequency might be even go to zero at a specific critical point, which signals the onset of charge density waves formation, as rationalized in 1D by the so-called Peierls instability \cite{peierls1955quantum}. This type of softening is profoundly different from the one described in the previous section as it is localized in a small region of finite wave-vector and does not appear at $k=0$. More in general, Kohn-like instabilities, defined as a localized decrease of the energy of acoustic phonons in a finite and small interval of wave-vectors, appear not only in association to CDW formation. 
They appear more generally whenever important nesting is exhibited by the Fermi surface, such as in NbC$_{1-x}$N$_{x}$ and NbN rocksalt structures \cite{Cohen_softening}. Intriguingly, similar softening mechanisms have been also reported in the acoustic dispersion relations of specific amorphous systems known as 'strain-glasses' \cite{ren2021boson}. 

Here, we want to assess the effects of Kohn-like instabilities in the metallic state of a superconducting material. The discussion in this subsection follows Ref.~\cite{jiang2022sharp}. Despite the main interest of this analysis is the question of coexistence and/or competition between CDW and superconductivity, we will leave the model as generic as possible in order to account for other possibilities not related to CDW. Moreover, for brevity, we will consider only the case of acoustic phonons, although the same qualitative behaviours hold for optical phonons as well. For more details and for the case of optical modes we refer to \cite{jiang2022sharp}.

As a phenomenological description of phonon softening, we consider the standard dispersion relation for acoustic phonons extracted from the denominator of the mediator Green's function in Eq. \ref{green} with
\begin{equation}\label{acu1}
    \Omega(k)=p(k)\left(v\,k-\frac{v}{2k_{\mathrm{VH}}}k^2\right)\,,\qquad \Gamma(k)=D k^2\,.
\end{equation}
In the equation above, $k_{\mathrm{VH}}$ is the wave-vector corresponding to the end of the Brillouin zone, the Van-Hove wave-vector. Most importantly, the softening of the dispersion is parameterized by the function $p(k)$ which is chosen as
\begin{equation}\label{soft1}
    p(k)=1-\zeta  \,\mathrm{exp}\left[-\left(\frac{k/k_\mathrm{VH}-\alpha}{\beta}\right)^2\right].
\end{equation}
The softening dip is assumed to be of Gaussian shape. The parameter $\zeta$ determines the depth of the softening, $\alpha$ controls the wave-vector at which the dip appears and $\beta$ its width. After assuming this dispersion, we can use the Eliashberg theory for damped bosonic mediators outlined in Section \ref{thel} to compute the various superconducting properties.

\begin{figure}
    \centering
    \includegraphics[width=0.8\linewidth]{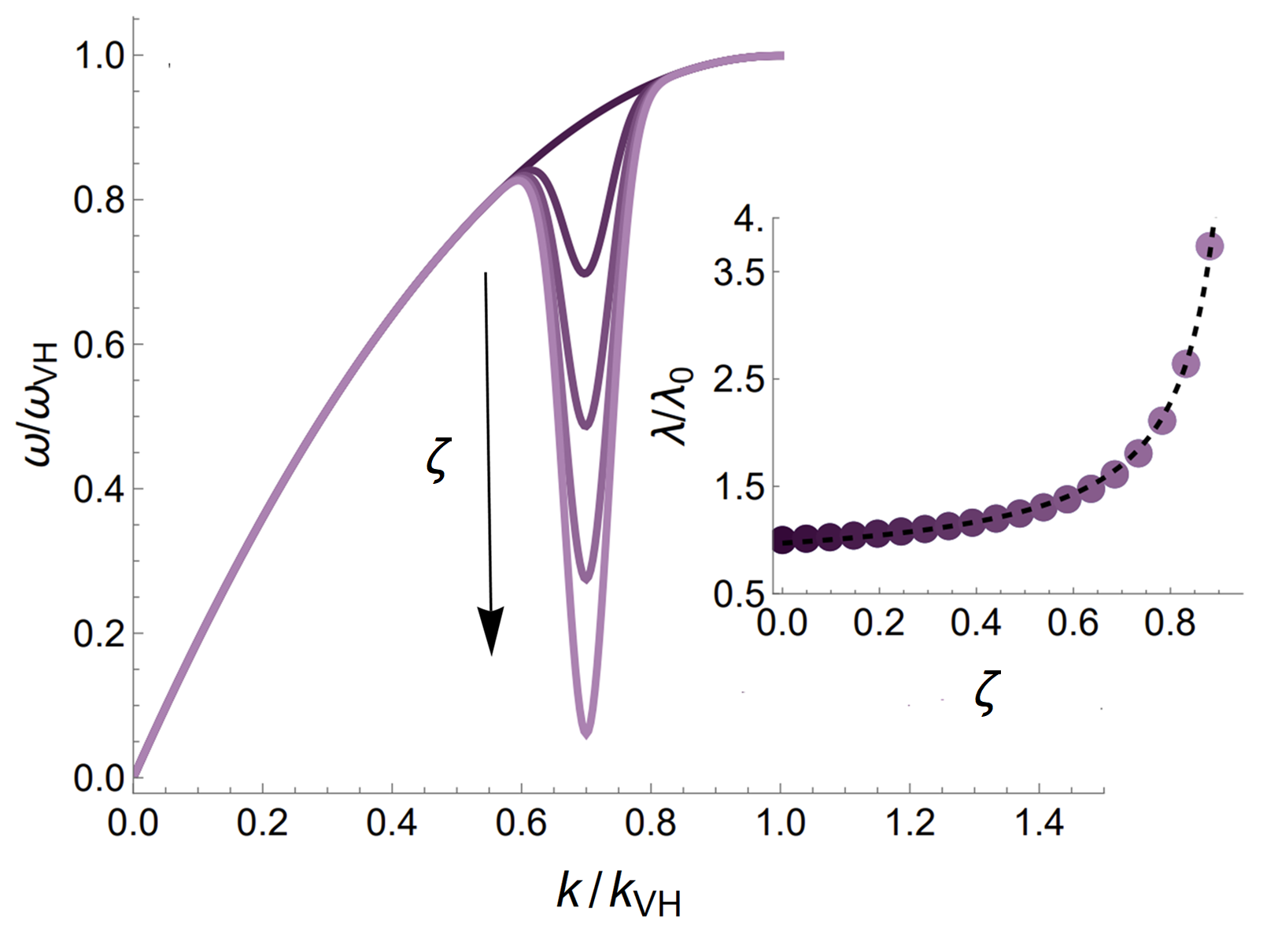}
    
    \vspace{0.1cm}
    
    \includegraphics[width=0.8\linewidth]{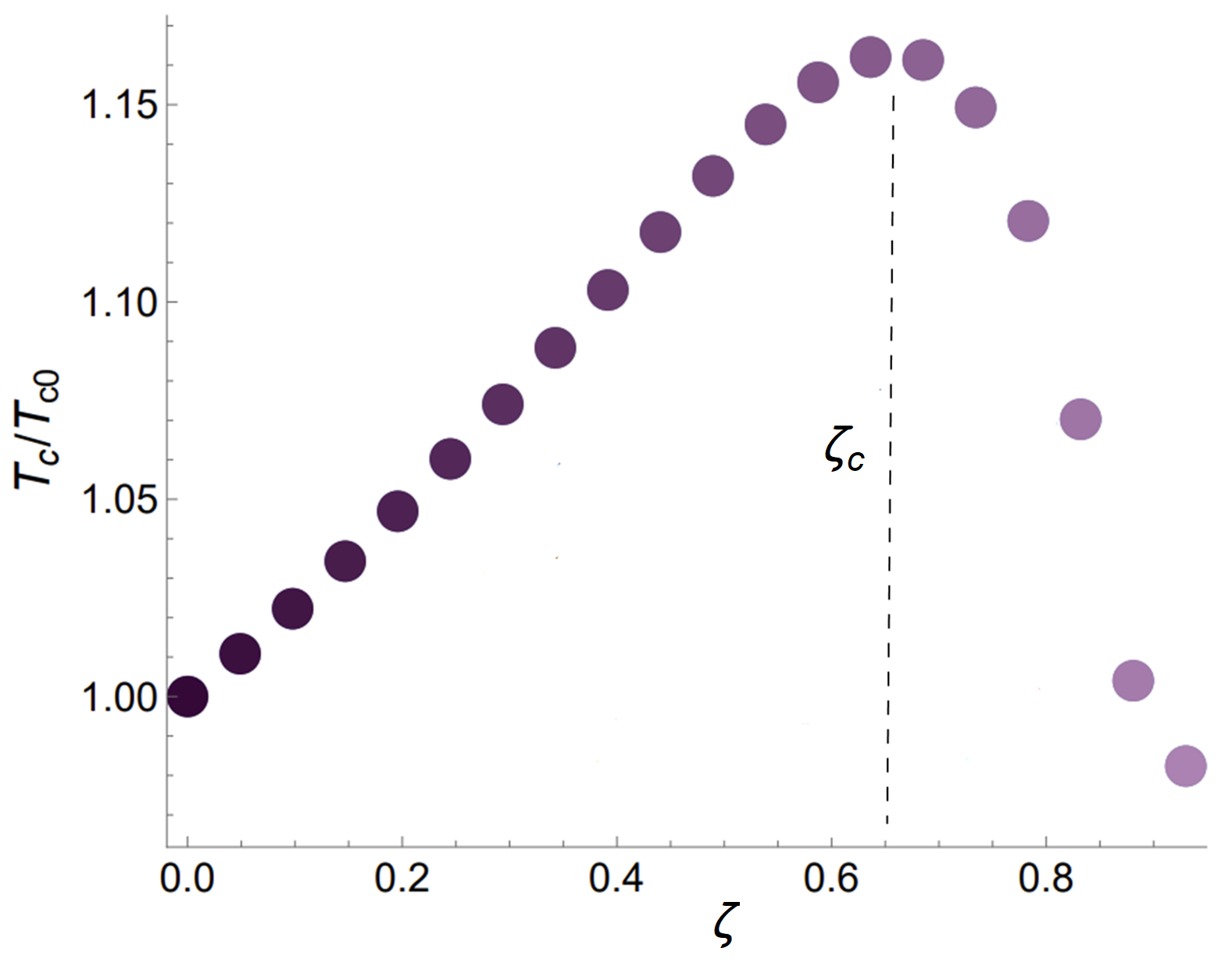}

    \vspace{0.1cm}
    
    \includegraphics[width=0.75\linewidth]{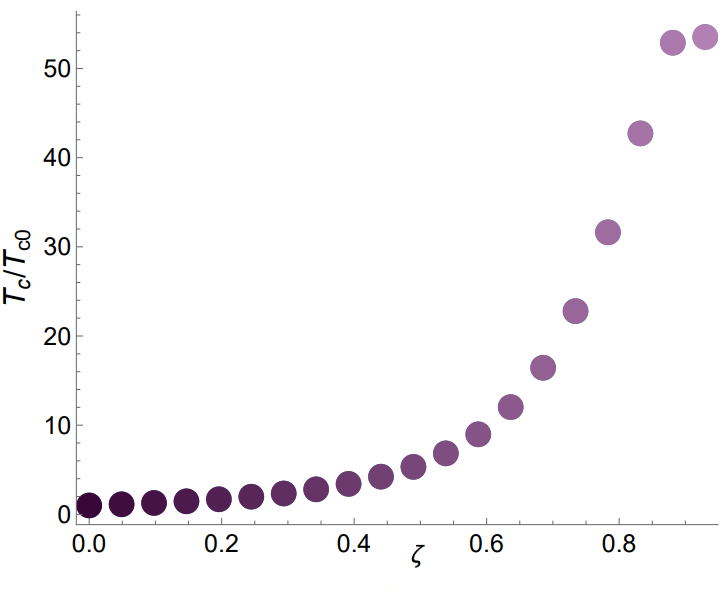}
    \caption{The effects of Kohn-like softening in the acoustic dispersion on the superconducting properties. \textbf{(Top)} the dispersion relation of the bosonic mediator for different values of the softening depth $\zeta$ and the corresponding behavior of the Eliashberg coupling $\lambda$. $\lambda_0$ stands for $\lambda(\zeta=0)$. \textbf{(Center and Bottom)} the critical temperature $T_c$ as a function of smaller and large softening depth $\zeta$ respectively. The vertical dashed line indicates the location of the optimal condition $\zeta=\zeta_c$. Copyright IOP from Ref.~\cite{jiang2022sharp}.}
    \label{fig:soft}
\end{figure}

In Fig.\ref{fig:soft}, we show the results as a function of the depth of the softening dip $\zeta$. Upon increasing the depth of the softening region, the value of the Eliashberg electron-phonon coupling $\lambda$ grows monotonically and it can be strongly enhanced. Nevertheless, this enhancement of the electron-phonon coupling is not always reflected in an increase of the superconducting temperature. On the contrary, a non-monotonic behavior is observed in the critical temperature, which first grows with softening roughly linearly, but then decreases quickly after a critical value $\zeta_c$. Intuitively, this non-monotonic trend is explained by the competition of the different factors appearing in the Allen-Dynes formula, Eq.\eqref{allenformula}. More precisely, despite the fact that the electron-phonon coupling $\lambda$ grows with softening, the logarithmic average frequency $\omega_{\text{log}}$ \eqref{eq13} decreases with it. As a consequence, a maximum value of $T_c$ appears at $\zeta=\zeta_c$. From a different perspective, the appearance of a maximum is compatible with the concept of ``optimal frequency'' developed by Bergmann and collaborators \cite{bergmann1973sensitivity}. In a nutshell, the idea is that a weight transfer in the Eliashberg function $\alpha^2 F(\omega)$, which in our case is induced by softening, is beneficial to superconductivity only when it is near the ``optimal frequency'', defined as the location in which the functional derivative $\frac{\delta T_c}{\delta \alpha^2 F(\omega)}$ is maximal. The optimal condition coincides with the maximum in $T_c$ at $\zeta_c$.

The value of $\zeta_c$ and the maximum increase in the critical temperature, $T_c(\zeta_c)/T_c(0)$, are strongly sensitive to the parameters of the model. Nevertheless, some general conclusions can be reached. In particular, in a weakly-coupled superconductor, the increase of $T_c$ due to softening can, in general, be very large (over one order of magnitude in $T_c$) but at the same time it requires a substantial degree of softening, \textit{i.e.}, a relatively large value of $\zeta$ (see example in the bottom panel of Fig.\ref{fig:soft}). On the contrary, for strongly-coupled systems, the increase of $T_c$ is more limited but the degree of softening needed to reach it is also smaller (see the central panel of Fig.\ref{fig:soft}). A more complete case study about the effects of Kohn-like softening on $T_c$, including the case of optical modes, can be found in \cite{jiang2022sharp}.

 \section{Applications to Emerging Quantum Materials}
 \subsection{Cuprates}
It may not have been a sheer coincidence that the major breakthrough in high-$T_c$ superconductivity, \textit{i.e.}, the Nobel-prize winning discovery of the cuprate rare-earth oxides by Bednorz and M\"{u}ller in 1986 \cite{Bednorz1986}, came in that same Zurich IBM lab after more than 15 years of studying the dielectric properties and soft-mode transitions in strontium titanate.
 Bednorz and M\"{u}ller's original intuition was that certain oxides could host Jahn-Teller type composites made of an electron plus a local lattice distortion that could travel as whole through the lattice, thus leading to a very strong electron-phonon coupling. 
 While lattice distortion and strong electron-phonon coupling have certainly been recognized to be important factors for the high $T_c$ of the cuprates, other non-trivial (\textit{e.g.} magnetic) phenomena have since also been observed, which also appear to strongly affect the $T_c$.

 Importantly, in a series of papers by Liarokapis, Kaldis and co-workers, Raman spectroscopy studies of the Cu-bonded oxygen atoms and associated Raman-active modes, highlighted a number of striking phonon-softening instabilities. The in-plane ($A_{g}$) oxygen vibrations in YBa$_2$Cu$_3$O$_x$ were shown in \cite{Liarokapis1} to suffer a major softening right at the optimal doping $x \approx 6.92$ that corresponds to the highest $T_{c}$. Concomitantly, a displacive structural phase transition involving the Cu$_2$O planes (basically a dimpling of the planes) was demonstrated in Ref. \cite{LiarokapisPRL} to also occur at a value of oxygen doping very close to the optimal one for $T_c$.
 
Recent experiments where the cuprate superconductor La$_{2-x}$Ba$_x$CuO$_4$ (LBCO) at $\frac{1}{8}$ doping was irradiated with protons~\cite{Welp2018} observed a (radiation) disorder induced enhancement of $T_c$ despite the proximate charge density wave (CDW) ordering temperature being unaffected by irradiation. The measurements found up to a 50\% increase in $T_c$ with the dosage of radiation. Above a critical value of the dosage,  $T_c$ was gradually suppressed until superconductivity was destroyed. To understand these observations, scenarios involving the competition between CDW and superconductivity seem promising at first sight, especially given their proximity in the phase diagram. However, given that the CDW transition temperature seems unaffected by irradiation, it is unlikely that a mechanism involving the competition between two mean field phases~\cite{Chubukov2012-Enhancement} is at play.  It is also unclear how scalar disorder affects two mean field phases in an asymmetric fashion without any parameter dependence, except under special circumstances~\cite{Chubukov2012-Enhancement, Hirschfeld2016} which may not hold for LBCO. Hence, to explain the non-monotonic $T_c$ dependence in LBCO as a function of irradiation, a mechanism that does not involve any competition between CDW and superconductivity is a potential candidate. 
 \par
 \begin{figure}[h!]
\includegraphics[width=0.9\linewidth]{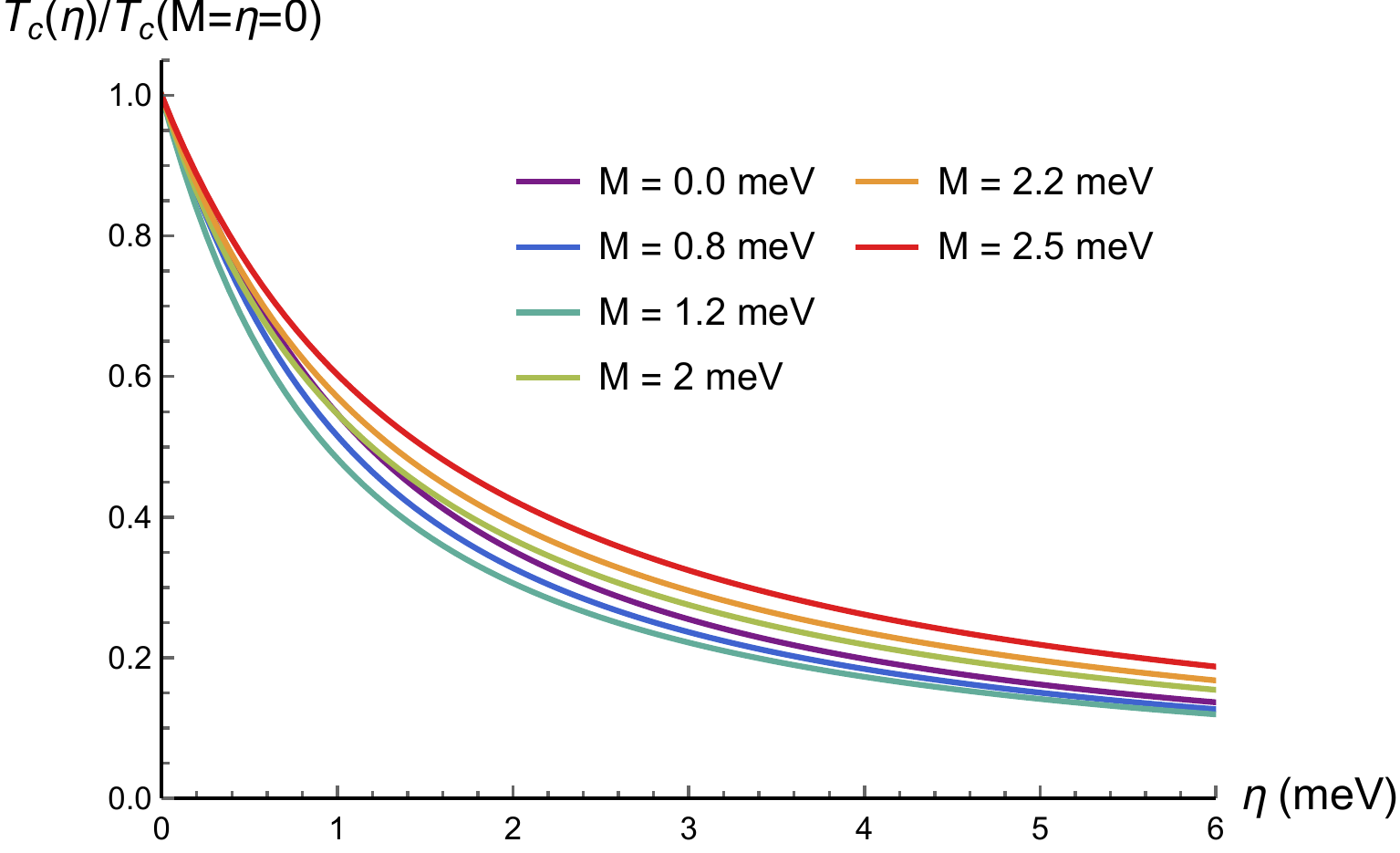}

\vspace{0.15cm}

\includegraphics[width=0.8\linewidth]{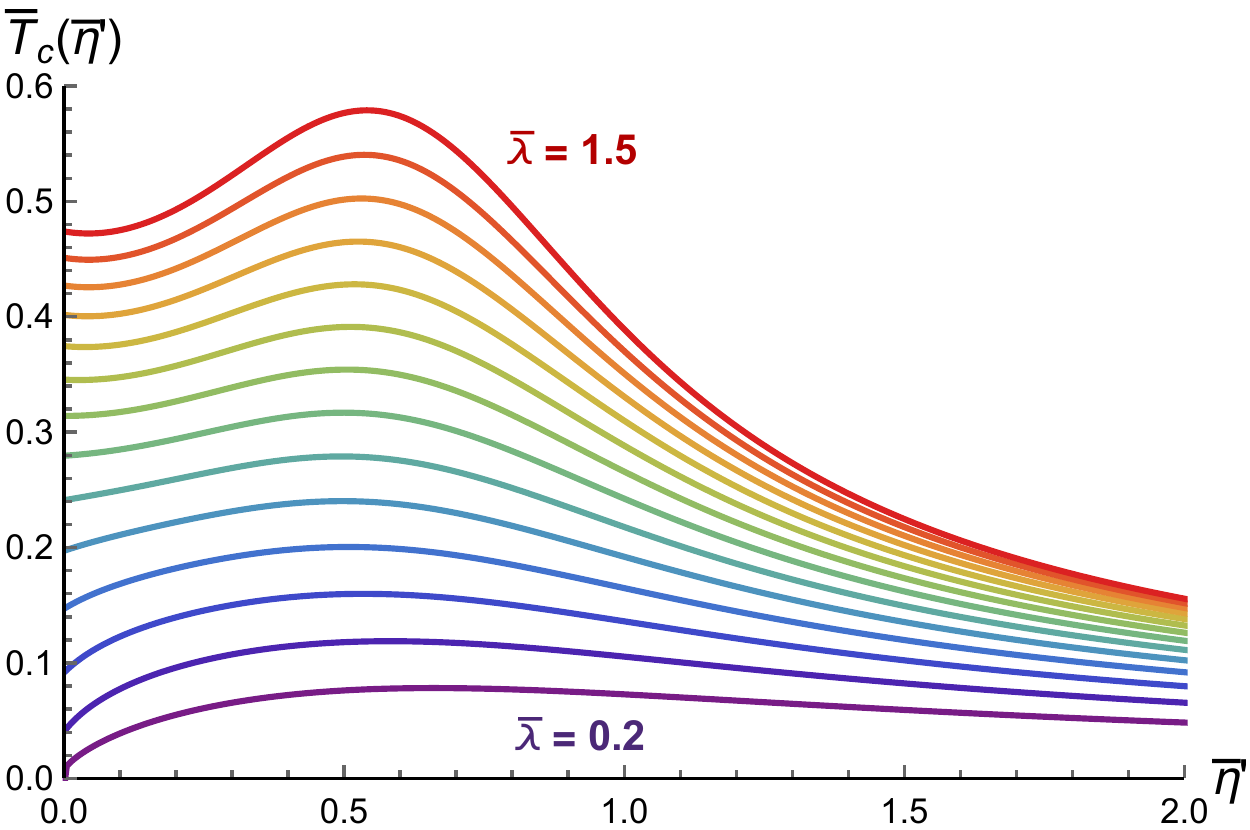}
\caption{\textbf{(Top)} Superconducting critical temperature $T_c$ (normalized to the case of zero dissipation and bosonic mass) as a function of the dissipation parameter $\eta$ for different masses $M$ for a local mediator ($\kappa=0$). \textbf{(Bottom)} For a non-local mediator ($\kappa \neq 0$): plot of the dimensionless $\bar{T}_c = T_c/\kappa$ as a function of $\bar{\eta}' = \eta'/\kappa$ for different dimensionless coupling strengths $\bar{\lambda} = \lambda/\kappa^2$ and $ M=0$. The $T_c$ peak is determined by $\kappa$, the energy scale arising from the bosonic spatial stiffness/velocity. 
} \label{Tc-Nonmonotonic}
\end{figure}
Ref.~\cite{Setty2019} made the case for enhanced $T_c$ due to glassy dissipation in spin fluctuation mediator. The SG phase has been observed in proximity to superconductivity in the cuprate~\cite{Aharony1995, Yamada2000, Ferretti2001, Imai2001, Julien2003, Birgeneau1990, Yamada2008, Hirota2018, Lin1999, Grafe2012, Julien2013} and iron superconductors~\cite{Curro2013, Mesot2013, Petrovic2015-Co, Petrovic2015-Ni, Buechner2012, Paulose2010}. It is thus reasonable to include the effects of the SG phase on the superconducting pairing.  Further there is plenty of direct experimental data~\cite{Julien2003, Aharony1995, Curro2013, Buechner2012, Lin1999, Curro2013, Imai2001, Mesot2013, Birgeneau1990, Julien2013} supporting a dissipative nature of the spin fluctuations mediating Cooper pairing. See Ref.~\cite{Setty2019} for a brief review of these experiments and their relevance to superconductivity in LBCO. From these discussions, the premise of a SG induced dissipative pairing mediator in LBCO has firm experimental support. We now follow Ref.~\cite{Setty2019} which argues that a non-local dissipative mediator can explain the proton irradiation  experiments in Ref.~\cite{Welp2018} despite the fact that the proximate charge density wave (CDW) transition is unaffected by disorder. \par

According to this proposal, disorder acts as an external tuning knob of the parameter $\eta$; hence, increased irradiation leads to larger dissipation in the pairing mediator. Then, for weak dissipation, $T_c$ rises and above a critical value of $\eta$ it gradually falls. To see this, the local ($\kappa =0$) and non-local ($\kappa \neq 0$) gap equations~\ref{Local-GapEqn}~\ref{Non-Local-GapEqn} can be solved for $T_c$ as a function of the dissipation parameter $\eta$.  Fig~\ref{Tc-Nonmonotonic} shows a plot of the solutions. For the case of a local mediator, the $T_c$ falls monotonically with dissipation. On the other hand, when the mediator is non-local,  $T_c$ is non-monotonic with dissipation parameter and reaches an optimal value at an $\eta$ value set by the stiffness. 
This can be understood from the energy integral leading to Eq.~\ref{Non-Local-GapEqn} above. For a non-local mediator, this integral  forces the gap equation to acquire dissipative contributions that both increase and decrease the effective coupling constant.  Note that weakly dissipative bosonic modes at different energy scales act coherently to enhance $T_c$ but eventually destroy superconductivity when dissipation dominates all the other energy scales.  Consequently, a non-monotonic behavior in $T_c$ follows. Thus the SG dissipative mechanism described above is a way to raise $T_c$ that does not rely on `tug-of-war' -like scenarios between two competing phases. Further analysis of the superconducting gap and specific heat as a function of the dissipation parameter can be found in Ref.~\cite{Setty2019}.

  \subsection{Hydrides}
Many recent studies have pointed to the possibility of achieving room-temperature superconductivity in the hydride compounds at high pressure \cite{Pickett}, following on Ashcroft's early intuition for metallic hydrogen \cite{Ashcroft}.
Recent experimental evidence points at superconductive behaviour in nitrogen-doped lutetium hydride thin films at pressures as low as 10kbar \cite{Ranga2023}.

While it is clear that phonon dispersion curves are strongly renormalized (\textit{i.e.}, lowered in energy) by anharmonicity in the hydrides \cite{Mauri2013,Mauri2014,Cohen_hydrogen}, a clear picture about the effect of the ubiquitous large anharmonicity on the superconductivity of these systems is missing. In particular, systematic studies of the anharmonic phonon linewidths and the effect thereof on the Cooper pairing are currently lacking. Since these systems exhibit high-T superconductivity at high pressures, the interplay between lattice dynamics under pressure, and anharmonicity, which leads to the resulting electron-phonon coupling, is expected to be non-trivial. In particular, the effects of pressure are twofold, on one hand there exists a critical pressure to stabilize the superconducting lattice structure \cite{Pickard2007,Pickett_2022,Zurek_review}, while on the other hand there are (hitherto much less explored) effects of pressure-mediated phonon dynamics on the pairing mechanism \cite{yesudhas2020origin,Setty2021}.

Recent progress \cite{Schrodi} has identified the phonon $E_{u}$ mode as the one mainly responsible for the pairing in atomic hydrogen at high pressure. The effect of phonon anharmonicity on the superconducting critical temperature $T_c$ has, instead, remained poorly understood. The anharmonic extension of BCS theory to include the effect of anharmonic damping on the pairing mechanism \cite{Setty2020}, has shown that anharmonicity can either enhance the $T_{c}$ or lower it, depending on the extent of phonon damping (moderate or very large, respectively), for the case of optical phonons, whereas for acoustic phonons the effect is always to cause a depression of the $T_c$.

This theory \cite{Setty2020} thus might explain why the $T_{c}$ is much lowered by huge anharmonicity of the low-lying optical phonons in aluminum \cite{Errea2021,Bergara2010}, palladium \cite{Mauri2014,Mauri2013}, and platinum hydrides \cite{Mauri2014}.

\begin{figure}
\includegraphics[width=0.8\linewidth]{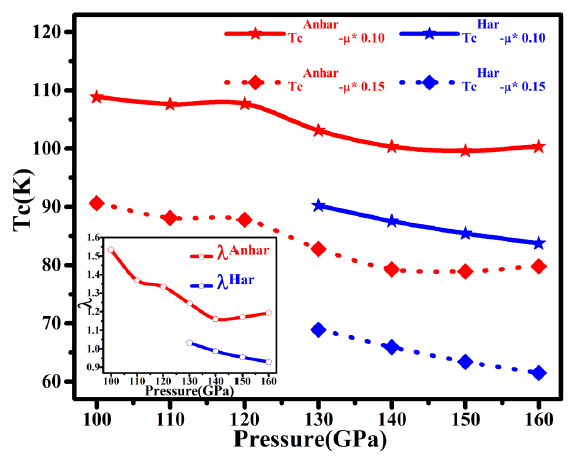}
\caption{Superconducting critical temperature $T_c$ and electron-phonon coupling constant $\lambda$ as a function of pressure computed with and without the anharmonic corrections for the $P6_{3}/mmc$ phase of ScH$_6$. Copyright AIP from Ref.~\cite{Errea2021}.}
\label{Errea}
\end{figure}

Conversely, an enhancement of the superconducting $T_c$, in a regime of moderate anharmonicity, may be responsible for the observed enhancement of $T_c$ due to anharmonicity in the high-pressure $P6_{3}/mmc$ phase of ScH$_{6}$ as reported in \cite{Errea2021}, see Fig. \ref{Errea}. Importantly, in the current literature, \textit{e.g.}, Ref.\cite{Errea2021}, the effects of anharmonicity are mostly considered at the level of the renormalization of the bare energy. So far, the effects of the anharmonic linewidth has been largely overlooked (see nevertheless \cite{2023arXiv230307962D} for a recent discussion about it).
In future studies, it could be useful to carry more systematic studies of the effect of phonon anharmonicity on the $T_c$ by more closely combining theoretical concepts \cite{Setty2020}, and atomistic computations \cite{EPW}.

  \subsection{The case of TlInTe$_2$}
  The discussion in this subsection closely follows Ref.~\cite{Setty2021}. TlInTe$_2$ undergoes a superconducting transition at a pressure of $5.7$ GPa with a T$_c \simeq$ $4$K~\cite{yesudhas2020origin}. The T$_c$ behaves non-monotonically with further increasing the pressure -- it decreases initially and climbs again with the minimum value of T$_c$ occurring at $10$ GPa. Concurrently, \textit{ab initio} electronic structure calculations found a Lifshitz transition induced change in Fermi surface topology between $6.5-9$ GPa and the formation of enlarged electron pockets at the Fermi level. Additionally,  X-ray diffraction and Raman scattering  measurements performed at  high pressure found that the $A_g$ phonon mode begins to soften~\cite{yesudhas2020origin}. Naively, the V-shaped T$_c$ behavior may be attributable to a combination of softening of the $A_g$ phonon mode and variations in the electronic density of states (DOS) with pressure. However, the T$_c$ appears to get reduced exactly in the regime where there is an increase in DOS from the electron pocket due to the Lifshitz transition. Moreover, the suppression of T$_c$ with increasing phonon frequency involves the Bergmann-Rainer criterion which would, in turn, require a second dip in $T_c$. This feature is, however, absent in the experimental observations. Hence, we can rule out a dominant role of  electronic DOS or phonon frequency shifts in understanding the observed T$_c$ dependence.   \par
  Here, we consider the role of both phonon frequency and linewidth in determining T$_c$ as a function of pressure in TlInTe$_2$.  Raman scattering linewidths extracted as a function of pressure indicate that anharmonicity in this material is in an optimal range --  weak enough so that the phonons remain coherent, but strong enough so as to have significant effects on the superconducting properties. In particular, as we will see below, T$_c$ correlates positively with the ratio of the linewidth to the peak frequency, $\Gamma/\omega_0$. The possibility of excluding other electronic DOS and phonon frequency-shift effects on the pairing renders TlInTe$_2$ an ideal playground to test the role of anharmonic boson damping. At this juncture, the properties of the normal state, strength of specific electron-phonon couplings, pairing symmetry etc are not completely determined in TlInTe$_2$. But the formalism and conclusions presented below are general enough so that the above uncertainties can be accommodated as more experimental data becomes available. 

  \begin{figure}[h]
    \centering
    \includegraphics[width=0.8\linewidth]{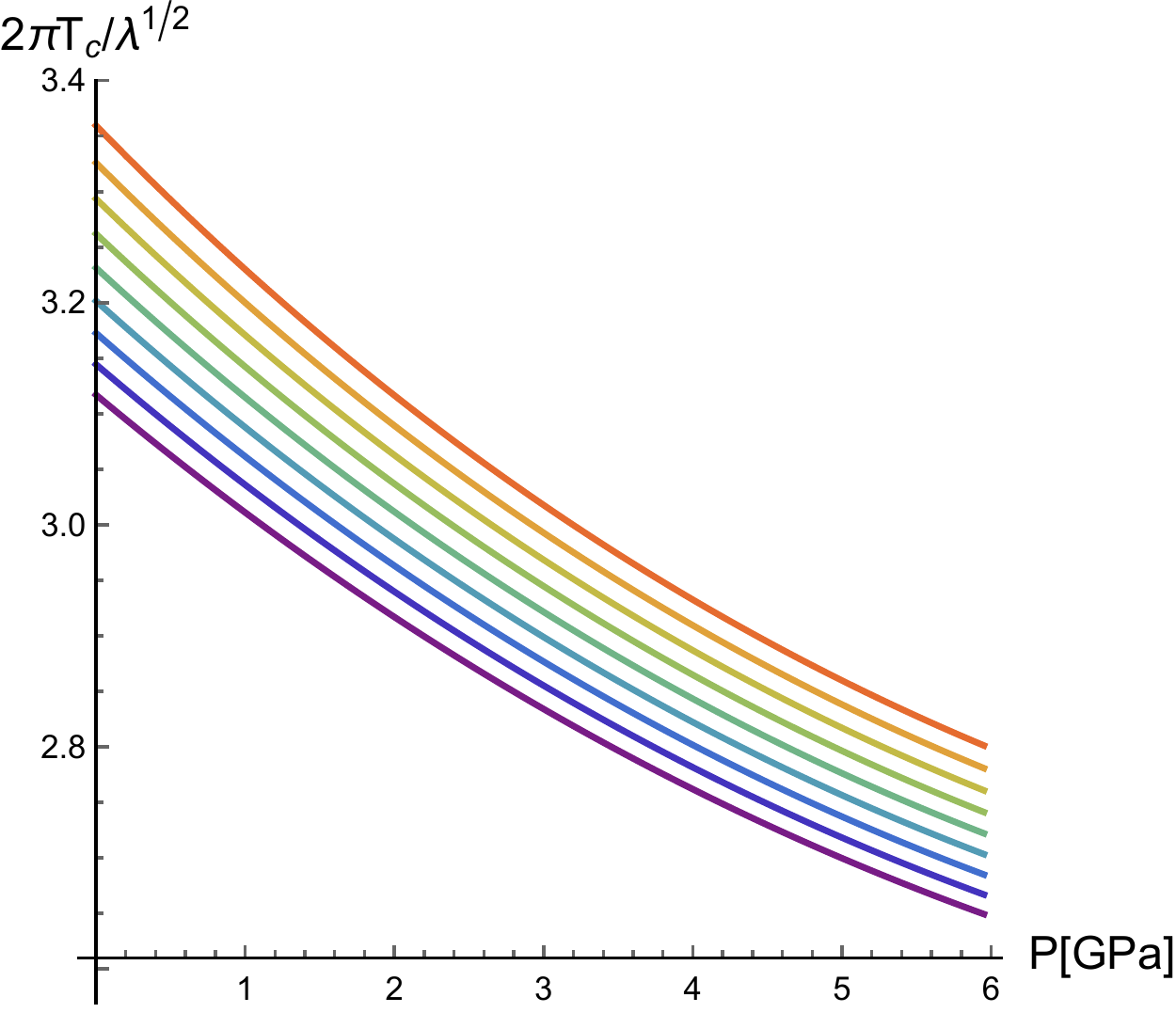}
    \caption{The normalized superconducting transition temperature as a function of pressure for various Klemens damping parameters $\Gamma=\alpha \omega_0^5$ where $\omega_0$ is the energy of the optical mode at zero wavevector. Here, $\alpha$ increases from $0.1$ to $1.0$ for red to purple curves. Copyright APS from Ref.~\cite{PhysRevB.106.139902}.
    }
    \label{fig2}
\end{figure}

  We begin by studying how the optical phonons of a crystal lattice are affected by external pressure. Predominantly, pressure acts to induce a volume contraction (negative volume change) in the material. We can relate change in volume to a change in phonon frequency through the Gr{\"u}neisen parameter, $\gamma = - d \ln \omega' / \ d \ln V$, via the relation~\cite{Kunc}:  
\begin{equation}
\frac{\omega'(V)}{\omega'_{P=0}}=\left(\frac{V}{V_0}\right)^{-\gamma}\,.
\label{Gruneisen}
\end{equation}
Here the optical phonon energy at zero ambient pressure is denoted by $\omega'_{P=0}$ and the relations above apply to individual phonon modes with frequency $\omega'$. 
The change in pressure can be written in terms of the volume change, as described by the Birch-Murnaghan equation of state \cite{Birch} (see also Sec.\ref{opop} above). The equation provides an expression for the pressure $P(V)$ and is derived based on nonlinear elasticity theory. 
We then replace $V$ with $\omega'$ in \eqref{Gruneisen} to obtain a relationship between applied pressure and optical phonon frequency $\omega'$~\cite{Kunc} given by
\begin{equation}
    P(X)\,=\,\frac{3}{2}\,b_0\,\left(X^{7}-X^{5}\right)\,\left[1+\eta\,(1-X^{2})\right]\label{kunc}\,,
\end{equation}
where we have defined $X \equiv (\omega'/\omega'_{P=0})^{1/3\gamma}$. Next, we invert the above Eq.\eqref{kunc} and obtain $\omega'$ as a function of $P$. We see that $\omega'$ is a function that monotonically increases with $P$ in the regime of interest, and is modulated by anharmonicity through $\gamma$. 
We have also defined $b_0 = B_0/\gamma_0$ where $B_0$ the bulk modulus and $\eta=(3/4)(4-B_{0}')$ with $B_{0}'=d B_{0}/dP$.
The frequency $\omega'$ refers to the real part of the phonon dispersion (including the renormalization shift from anharmonicity~\cite{Baowen}). The imaginary part of the dispersion relation is related to the phonon damping coefficient $\Gamma$ (the inverse phonon lifetime). These quantities are given by the following relations (see for example Eqs. (23)-(27) in Ref.~\cite{Baowen})
\begin{eqnarray}
    \omega^2&\,=\,&\omega_0^2\,-\,i\,\omega\,\Gamma\,+\,\mathcal{O}(q^2),\\ 
    \omega' &\equiv& \mathrm{Re}(\omega)\,=\,\frac{1}{2}\,\sqrt{4\,\omega_0^2\,-\,\Gamma^2}\,+\mathcal{O}(q^2),\quad \\
    \frac{\Gamma}{2}&\,\equiv\,&\mathrm{Im}(\omega)\,+\mathcal{O}(q^2).
\label{optical}
\end{eqnarray}
The phonon linewidth $\Gamma$ can, in principle, be evaluated from quantitative microscopics using the Self-Consistent Phonon (SCP) methodology~\cite{Baowen,Tadano1} for specific systems~\cite{Tadano2}.  Here we rather focus on generic qualitative trends in terms of the effect of $\Gamma$ on the pairing and on $T_c$. 

\begin{figure}[t]
    \centering
     \includegraphics[width=0.8\linewidth]{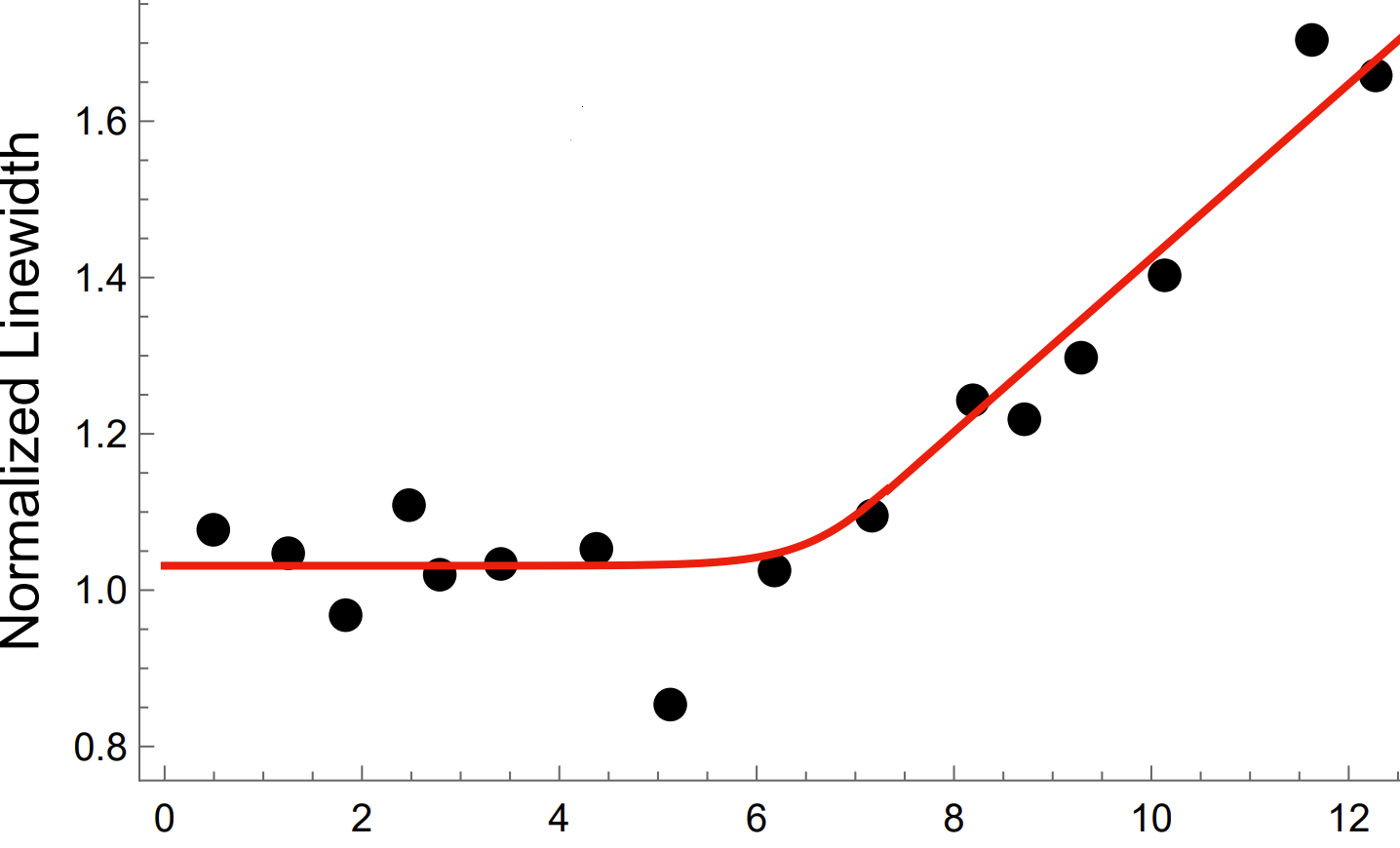}
     
     \vspace{0.2cm}
     
      \includegraphics[width=0.8\linewidth]{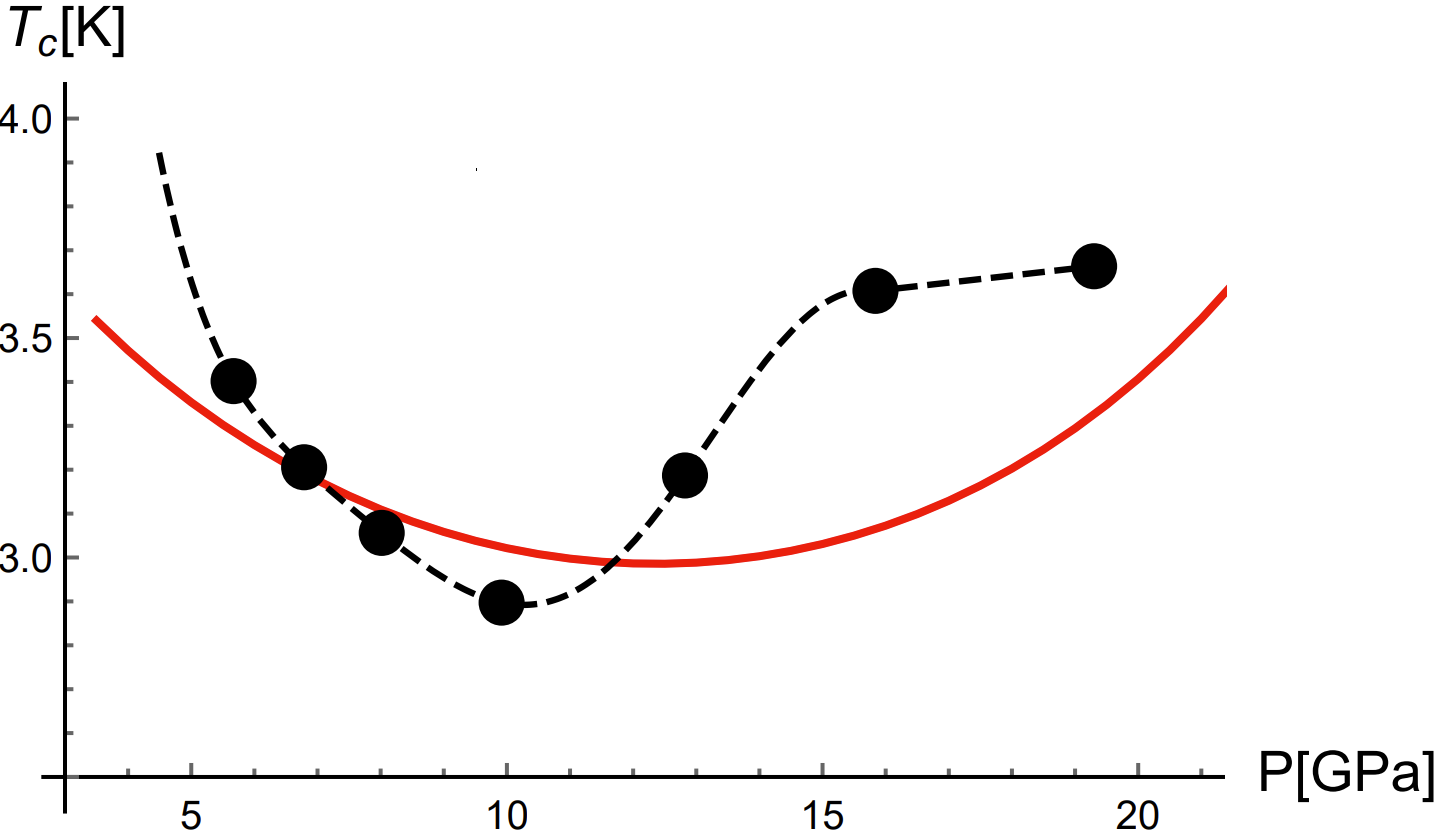}
     
    \caption{\textbf{(Top)} Phonon linewidth (black symbols) extracted from experimental data \cite{yesudhas2020origin} versus optimal phenomenological fit (red solid line). \textbf{(Bottom)} Comparison between the theoretically obtained calculations for the critical temperature (red solid line) and experimental data (black symbols). The dashed black line is an interpolation of the experimental data as a visual guide. For details about the parameters and the numerical procedure see \cite{PhysRevB.106.139902}. Copyright APS from Ref.~\cite{PhysRevB.106.139902}.}
    \label{TIT}
\end{figure}

As a concrete application of this model, we consider the case of TlInTe$_2$ using the data reported n Ref.~\cite{yesudhas2020origin}. We fit the bare frequency $\omega_0$ and the linewidth $\Gamma$ of the optical mode A$_g$, which is the dominant one in the electron pairing, as a function of the pressure $P$. The results for the linewidth are shown in the top panel of Fig.\ref{TIT}. We then use these functions as an input into the theoretical gap-equation framework and predict the corresponding $T_c$. First, we use the fitting for the energy $\omega_0(P)$ together with the Klemens expression for the linewidth $\Gamma= \alpha \omega_0^5$, with $\alpha$ a phenomenological parameter. The results for different values of $\alpha$ are shown in Fig.\ref{fig2}. The critical temperature $T_c$ decreases monotonically with the pressure. An approximately linearly decaying trend of $T_c$ with $P$ has been recently observed in the strongly anharmonic AlH$_{3}$ high-pressure hydride~\cite{Errea2021} as well as in the SC-I phase of CeH$_{10}$ in Ref.~\cite{Oganov}.
In more standard systems, a linear decay of $T_c$ with increasing $P$ has been reported in the literature for simple (\textit{e.g.}, elemental) superconductors~\cite{Ginzburg,Palmy,Chu}. In order to improve the results, we also used the fitted linewidth from the experimental data, available in \cite{yesudhas2020origin}for TlInTe$_2$. The predicted critical temperature is compared with the experimental data in the bottom panel of Fig.\ref{TIT}. The agreement, at least at qualitative level, is good. In particular, the theoretical prediction shows a minimum in the critical temperature at a pressure which roughly corresponds to the position of the minimum in $\Gamma/\omega_0$. This is rationalized, within the theoretical framework, by noticing that in the so-called coherent (moderate-damping) regime, where $\Gamma/\omega_0 \ll 1$, the behavior of $T_c$ positively correlates with the ratio $\Gamma/\omega_0$. Similar considerations also appear to hold for the Osmium based pyrochlore superconductors where T$_c$ shows an optimum value as a function of anharmonicity parameters~\cite{Hattori2012, Tsunetsugu2010}. 

  \subsection{SrTiO$_3$ and BaTiO$_3$}\label{pala}
  The discussions in this subsection follow from Ref.~\cite{Setty2022}. Superconductivity in the quantum paraelectric semiconductor SrTiO$_3$ has recently attracted much attention in view of different experimental protocols that have been discovered in order to promote superconductivity, often via a superconducting ``dome'', in the vicinity of the ferroelectric instability. These methods include doping in terms of carrier concentration, isotopic doping, mechanical strain, and dislocations. 
  Since the pioneering experimental work of M\"{u}ller and co-workers \cite{Mueller}, this material has been classified as a quantum paraeletric, in the sense that although a ferroelectric instability appears to be approached upon lowering the temperature below about $30$K, eventually the TO mode energy remains finite and real, and no condensation of the TO mode into the ferroelectric phase occurs (hence no real soft mode instability occurs). The common explanation for this phenomenon is that large quantum fluctuations of the lattice (in particular, zero-point motions of the oxygen atoms) prevent the mode condensation and the corresponding ``freezing'' of atomic position into polar order. Hence the ferroelectric transition is \emph{de facto} suppressed and the expected Curie-Weiss behaviour of the dielectric constant is instead replaced by a low-$T$ plateau \cite{Mueller}, hence the quantum paraelectric phase at $T < 4$K.  
  Mechanical strain has been shown to be a most effective way of re-establishing ferroelectricity in SrTiO$_{3}$, to the point that even ferroelectricity at room temperature and above has been demonstrated for SrTiO$_{3}$ material under strain. 
  
  Also, electron doping and oxygen doping have proved to be effective ways of inducing the ferroelectric instability and thus stabilize the ferroelectric phase. 
  The mechanism by which superconductivity occurs in this material has been thought for a long time to be puzzling because SrTiO$_{3}$ behaves as a superconductor even at very low carrier doping levels. Early evidence however has been collected pointing to the fact that it is indeed the soft transverse optical (TO) phonons which mediate the electron pairing \cite{Appel1969} in doped SrTiO$_{3}$, a mechanism recently confirmed \cite{Lonzarich2020}. More recently DFT calculations have demonstrated that indeed the maximum of the dome observed in the superconducting $T_c$ in SrTiO$_{3}$ does coincide with the TO mode energy crossing zero, \textit{i.e.}, with the ferroelectric criticality \cite{Spaldin}. This has become widely known as the ``quantum criticality''  paradigm for superconductivity in SrTiO$_{3}$, in view of the fact that large lattice fluctuations upon approaching the ferroelectric transition are thought to promote the Cooper pairing, and these fluctuations at such low temperatures are quantum, since the oxygens motion is of zero-point type. 
  
  In reality, however, these are just large atomic fluctuations about the equilibrium positions in the lattice, and quantum or not, they are always associated with large anharmonicity, simply because atoms displaced far away from the harmonic bonding minimum locally experience a large anharmonic potential, thus leading to huge values of the cubic derivative of the potential $V'''(a)$ in \eqref{AtomicPotential}, and hence to large values of the Gr\"{u}neisen parameter. This is indeed what happens, and giant values of Gr\"{u}neisen parameter $\gamma$ have been indeed observed both numerically and experimentally \cite{Balatsky}. This picture of giant anharmonicity assisting superconductivity even at low carrier concentrations in SrTiO$_3$ is further corroborated by recent anharmonic phonon calculations \cite{DelaireSrTiO3}.
  
  Finally, a similar dome in the superconducting $T_c$ with a maximum coinciding with the ferroelectric transition at which the TO mode goes to zero has been observed in the standard ferroelectric compound BaTiO$_3$. Also in this case very large anharmonicity of the TO phonon which explodes towards the ferroelectric transition, accompanies the superconducting dome.
  The fact that a very similar phenomenology is shared by quantum paraelectric SrTiO$_3$ and standard ferroelectric BaTiO$_3$ thus strongly suggests that the so-called quantum criticality may not be the peculiarity behind the superconducting properties of SrTiO$_3$, and points rather a unifying common mechanism.
  
  A recent proposal is that anharmonicity in SrTiO$_3$ leads to an even stronger pairing tendency, according to the anharmonic damping enhancement of $T_c$ mechanism proposed in \cite{Setty2020} and \cite{Setty2022} (see section \ref{secsoft} for details), which therefore explains the surprising fact that superconductivity in SrTiO$_3$ occurs at very low electron doping. Cfr. Fig.\ref{fig:boh} in Sec.\ref{secsoft} where the experimentally observed dome in $T_c$ is reproduced in model calculations that fully take into account the increase in the phonon anharmonic linewidth upon approaching the ferroelectric transition on both sides \cite{Setty2022}. But the general mechanism, by which superconductivity is enhanced by phonon damping is the same which is operative in standard (non-quantum) ferroelectrics such as BaTiO$_3$.

 \section{Outlook and Conclusions}
 \subsection{Open issues}
Several questions regarding the broader implications of bosonic damping effects on superconductivity remain pertinent. These questions range from material specific aspects that require a generic design principle of superconducting materials where boson anharmonic decoherence is a key player, to more fundamental aspects that may require new theoretical frameworks to deal with the role of dissipation on the superconducting ground state.

 In regards to the former, there is an immediate need to integrate first principles evaluation of phonon/bosonic linewidths into routines (such as Electron-Phonon Wannier (EPW)~\cite{EPW}) that evaluate material-specific superconducting properties. This could help make predictions about the role of bosonic anharmonic decay on Eliashberg functions, superconducting coupling constants, and critical temperatures.  Moreover, studying the role of multiple phononic branches and symmetry allowed dissimilarities in their anharmonicities can help make a clearer connection to real material systems. It is our hope that this review puts together various relatively disconnected theoretical, numerical and experimental works on the topic of anharmonic decoherence and superconductivity, and motivates future numerical work in this direction. 

On the formalism side, there are several open questions that remain unanswered. The simplest extension to the results presented would incorporate  the full electronic and bosonic self-energies, including anharmonic decoherence effects, into the Eliashberg formalism~\cite{Marsiglio2020}. This would give a  better understanding on the robustness of the conventional BCS results when retardation effects are taken into consideration.  Moreover, in the presence of a lattice, discrete spatial symmetries constrain the momentum dependence of higher order anharmonic terms in the Hamiltonian. The role of these symmetries and its interplay with decoherence and superconductivity is an interesting question worth exploring. Further, whether anharmonic damping can be a primary driver of superconductivity instead of playing a catalyst is also unclear.  Going beyond infinitesimal weak coupling effects is another promising avenue. This includes role of anharmonic damping on BCS-BEC (Bose-Einstein Condensation) crossover physics~\cite{Zwerger2008} and strong coupling superconductivity. Additionally, at intermediate coupling scales and anisotropic interactions, the superconducting ground state is modulated~\cite{Setty2022-PDW1, Setty2021-PDW2}; the role of anharmonicity on such ground states could have consequences on their stability. A more ambitious goal would be to include electron-electron correlation effects and understand their co-action with anharmonicity and superconductivity.  While this might seem like a difficult endeavor, recent progress has been made~\cite{Setty2020-LS, Setty2021-LS, Phillips2020} in obtaining exact solutions to the pairing problem in the presence of long range interactions. Therefore, gaining an intuitive understanding of the cooperation between anharmonic damping, strong correlations and superconductivity is a realistic possibility. 
 \subsection{Summary}
 In this review, we have addressed the role of anharmonic decoherence/damping effects on superconducting pairing properties using minimal theoretical models. We have adopted a mechanistic approach to describe the physics of how superconducting properties such as Eliashberg functions, coupling strengths and transition temperatures are affected, and applied these mechanisms to phenomenologically describe experiments on a variety of emerging quantum materials. The central theme of the review emphasizes the qualitative role played by damping, and the use of simplifying assumptions, rather than an elaborate implementation of atomistic first principle simulations. However, our objective is to highlight these simple mechanistic effects to motivate further works that could combine first principles evaluation of bosonic lineshapes with superconductivity routines such as EPW for material specific results. To this end, this review provided a basic introduction to phonon anharmonicity and various bosonic damping mechanisms that may be relevant to superconductivity. We then reviewed several experimental probes that can be used to measure anharmonic damping effects in pairing mediators followed by a brief interlude into existing literature that directly calculates phonon linewidths from first principles. The bulk of the remainder of the paper focused on minimal theories of superconductivity driven by damped bosons including phonons and glassy spin fluctuations models, and how they can be relevant to emerging quantum materials. As an outlook, we presented several outstanding problems and natural extensions of the current work that remain currently unaddressed. While a complete picture that delineates the role of anharmonicity and damping effects on superconductivity remains elusive, we believe that its fuller understanding holds the potential for interesting fundamental physics, novel numerical implementations, and innovative design of quantum materials and experimental realizations. 
 In particular, we anticipate broad implications of the anharmonic phyiscs of superconductivity also in nanostructured devices \cite{Fomin,Travaglino} and disordered materials \cite{Zacconebook}.
 
 \subsection*{Acknowledgements}
 M.B. acknowledges the support of the Shanghai Municipal Science and Technology Major Project (Grant No.2019SHZDZX01) and the sponsorship from the Yangyang Development Fund. A.Z. gratefully acknowledges funding from the European Union through Horizon Europe ERC Grant number: 101043968 ``Multimech'', and from US Army Research Office through contract nr.W911NF-22-2-0256.
 \bibliographystyle{apsrev4-1}
\bibliography{AnharmonicityReview.bib}
\end{document}